\documentclass[12pt,journal,draftcls,a4paper,onecolumn]{IEEEtran}

\usepackage{dsfont}
\usepackage{cleveref}
\usepackage{amssymb}
\usepackage{amsfonts}
\usepackage{graphicx}
\usepackage{amsmath}
\usepackage[all,poly]{xy}
\usepackage{multirow}
\usepackage{algorithm}
\usepackage{algorithmic}
\usepackage{color}
\usepackage[noadjust]{cite}

\newcommand{\bg}{\boldsymbol{g}}

\newcommand{\bh}{\boldsymbol{h}}
\newcommand{\bm}{\mathbf{m}}

\newcommand{\bp}{\boldsymbol{p}}
\newcommand{\bq}{\boldsymbol{q}}

\newcommand{\btt}{\boldsymbol{t}}

\newcommand{\bx}{\boldsymbol{x}}

\newcommand{\bX}{{\boldsymbol X}}

\newcommand{\bPhi}{{\boldsymbol \Phi}}
\newcommand{\bsigma}{{\boldsymbol \sigma}}
\newcommand{\bSigma}{{\boldsymbol \Sigma}}

\newcommand{\blue}{\textcolor{blue} }
\graphicspath{{figures/}}

\def\bfs{{\mathbf{s}}}

\def\calG{{\mathcal{G}}}

\def\calI{{\mathcal{I}}}

\newcommand{\Vpix}[1]{\mathbf{y}_{#1}}

\newcommand{\MATpix}{\mathbf{Y}}
\newcommand{\pix}[2]{y_{#1,#2}}

\newcommand{\Vpixels}{\mathbf{y}}

\newcommand{\nbpix}{N}
\newcommand{\nopix}{n}

\newcommand{\nbband}{L}

\newcommand{\nbmat}{R}
\newcommand{\nomat}{r}

\newcommand{\MATmat}{{\mathbf M}}
\newcommand{\Vmat}[1]{{\mathbf m}_{#1}}
\newcommand{\mat}[2]{m_{#1,#2}}

\newcommand{\MATabond}{{\bold A}}

\newcommand{\abond}[2]{{a}_{#1,#2}}
\newcommand{\Vabond}[1]{{\boldsymbol{a}}_{#1}}
\newcommand{\abonds}[1]{a_{#1}}
\newcommand{\Vabonds}{{\boldsymbol{a}}}

\newcommand{\Vnonlin}{\boldsymbol{b}}
\newcommand{\nonlin}[1]{b_{#1}}

\newcommand{\Vtabond}[1]{\boldsymbol{z}_{#1}}

\newcommand{\tabond}[2]{z_{#1,#2}}
\newcommand{\MATtabond}{{\bold Z}}

\newcommand{\Vnoise}{{\mathbf e}}

\newcommand{\MATnoise}{{\bold E}}

\newcommand{\noisevar}{\sigma^2}

\newcommand{\paramvect}{\boldsymbol{\theta}}

\newcommand{\Simplex}{\mathcal{S}}

\newcommand{\R}{\mathds{R}}

\newcommand{\transp}{^T}

\newcommand{\etr}{\mathrm{etr}}

\newcommand{\diag}[1]{\textrm{diag}\left(#1\right)}
\newcommand{\Ndistr}[1]{\mathcal{N}\left(#1\right)}

\newcommand{\norm}[1]{\left\|#1\right\|}

\newcommand{\Vun}[1]{{\boldsymbol{1}}_{#1}}
\newcommand{\Vzero}{\boldsymbol{0}}
\newcommand{\Id}[1]{\textbf{I}_{#1}}
\newcommand{\Indicfun}[2]{\textbf{1}_{#1}\left(#2\right)}

\newenvironment{algogo}[1]{
\smallskip
\noindent \hrule\vspace{0.2\baselineskip} \hrule
\begin{small}
\refstepcounter{algo} \center{\bf \textsc{Algorithm \thealgo}}
\\{\center{\bf #1}}
\smallskip
\flushleft
 } {
\end{small}
\smallskip
\hrule\vspace{0.2\baselineskip} \hrule
\smallskip
}

\newcounter{algo}
\renewcommand{\thealgo}{\arabic{algo}}

\title{Unsupervised Post-Nonlinear Unmixing of\\ Hyperspectral Images Using\\
a Hamiltonian Monte Carlo Algorithm}

\author{Yoann Altmann, Nicolas Dobigeon and Jean-Yves Tourneret
\thanks{Part of this work has been supported by Direction Generale de l'armement, French Ministry of defence, and by the Hypanema ANR
Project ANR Project n$^\circ$ANR- 12-BS03-003.}
\thanks{The authors are with the University of Toulouse, IRIT/INP-ENSEEIHT/T\'eSA, 2 rue Charles Camichel, BP 7122, 31071 Toulouse cedex 7,
France (e-mail: \{Yoann.Altmann, Nicolas.Dobigeon,
Jean-Yves.Tourneret\}@enseeiht.fr).}}

\begin{document}
 \maketitle

\begin{abstract}
This paper presents a nonlinear mixing model for hyperspectral image
unmixing. The proposed model assumes that the pixel reflectances are
post-nonlinear functions of unknown pure spectral components
contaminated by an additive white Gaussian noise. These nonlinear
functions are approximated using polynomials leading to a polynomial
post-nonlinear mixing model. A Bayesian algorithm is proposed to
estimate the parameters involved in the model yielding an
unsupervised nonlinear unmixing algorithm. Due to the large number
of parameters to be estimated, an efficient Hamiltonian Monte Carlo
algorithm is investigated. The classical leapfrog steps of this
algorithm are modified to handle the parameter constraints. The
performance of the unmixing strategy, including convergence and
parameter tuning, is first evaluated on synthetic data. Simulations
conducted with real data finally show the accuracy of the proposed
unmixing strategy for the analysis of hyperspectral images.
\end{abstract}

\newpage

\section{Introduction}
\label{sec:intro} Identifying macroscopic materials and quantifying
the proportions of these materials are major issues when analyzing
hyperspectral images. This blind source separation problem, also
referred to as spectral unmixing (SU), has been widely studied for
the applications where the pixel reflectances are linear
combinations of pure component spectra
\cite{Craig1994,Heinz2001,Eches2010a,Miao2007a,Yang2011}. However,
as explained in \cite{Keshava2002,Bioucas2012}, the linear mixing
model (LMM) can be inappropriate for some hyperspectral images, such
as those containing sand, trees or vegetation areas. Nonlinear
mixing models (NLMMs) provide an interesting alternative for
overcoming the inherent limitations of the LMM. They have been
proposed in the hyperspectral image literature and can be divided
into two main classes.

 The first class of NLMMs consists of physical
models based on the nature of the environment. These models include
the bidirectional reflectance based model proposed in
\cite{Hapke1981} for intimate mixtures associated with sand-like
materials and the bilinear models recently studied in
\cite{Somers2009,Nascimento2009,Fan2009,Halimi2010} to account for
scattering effects mainly observed in vegetation and urban areas.
The second class of NLMMs contains more flexible models allowing
different kinds of nonlinearities to be approximated. These flexible
models are constructed from neural networks
\cite{Guilfoyle2001,Altmann2011IGARSS}, kernels
\cite{Chen2012,Altmann2013}, or post-nonlinear transformations
\cite{Jutten2003,Babaie2001}. In particular, a polynomial
post-nonlinear mixing model (PPNMM) has recently shown interesting
properties for the SU of hyperspectral images \cite{Altmann2012}.

Most nonlinear unmixing strategies available in the literature are
supervised, i.e., the endmembers contained in the image are assumed
to be known (chosen from a spectral library or extracted from the
data by an endmember extraction algorithm (EEA)). Moreover, most
existing EEAs rely on the LMM
\cite{Winter1999,Nascimento2005,Chaudhry2006} and thus can be
inaccurate for nonlinear mixtures. Recently, a nonlinear EEA based
on the approximation of geodesic distances has been proposed in
\cite{Heylen2011} to extract endmembers from the data. However, this
algorithm can suffer from the absence of pure pixels in the image
(as most linear EEAs).

This paper presents a fully unsupervised Bayesian unmixing algorithm
based on the PPNMM studied in \cite{Altmann2012}. In the Bayesian
framework, appropriate prior distributions are chosen for the
unknown PPNMM parameters, i.e., the endmembers, the mixing
coefficients, the nonlinearity parameters and the noise variance.
The joint posterior distribution of these parameters is then
derived. However, the classical Bayesian estimators cannot be easily
computed from this joint posterior. To alleviate this problem, a
Markov chain Monte Carlo (MCMC) method is used to generate samples
according to the posterior of interest. More precisely, due to the
large number of parameters to be estimated we propose to use a
Hamiltonian Monte Carlo (HMC) \cite{Duane1987} method to sample
according to some conditional distributions associated with the
posterior. HMCs are powerful simulation strategies based on
Hamiltonian dynamics which can improve the convergence and mixing
properties of classical MCMC methods (such as the Gibbs sampler and
the Metropolis-Hastings algorithm) \cite{Brooks2011,Robert2004}.
These methods have received growing interest in many applications,
especially when the number of parameters to be estimated is large
\cite{Neal1996,schmidt09funcfact}. The classical HMC can only be
used for unconstrained variables. However, new HMC methods have been
recently proposed to handle constrained variables \cite[Chap.
5]{Brooks2011} \cite{Hartmann2005,Brubaker2012} which allow HMCs to
sample according to the posterior of the Bayesian model proposed for
SU. Finally, as in any MCMC method, the generated samples are used
to compute Bayesian estimators as well as measures of uncertainties
such as confidence intervals.

The paper is organized as follows. Section \ref{sec:Problem}
introduces the PPNMM for hyperspectral image analysis. Section
\ref{sec:bayesian} presents the hierarchical Bayesian model
associated with the proposed PPNMM and its posterior distribution.
The constrained HMC (CHMC) algorithm used to sample some parameters
of this posterior is described in Section \ref{sec:CHMC}. The CHMC
is coupled with a standard Gibbs sampler presented in Section
\ref{sec:Gibbs}. Some simulation results conducted on synthetic and
real data are shown and discussed in Sections \ref{sec:simu_synth}
and \ref{sec:simu_real}. Conclusions are finally reported in Section
\ref{sec:conclusion}.

\section{Problem formulation}
\label{sec:Problem}
\subsection{Polynomial post-nonlinear mixing model}
\label{subsec:PPNMM} This section recalls the nonlinear mixing model
used in \cite{Altmann2012} for hyperspectral image SU. We consider a
set of $N$ observed spectra $\Vpix{n} =
[\pix{n}{1},\ldots,\pix{n}{\nbband}]\transp, n \in \left \lbrace
1,\ldots,N \right \rbrace$ where $\nbband$ is the number of spectral
bands. Each of these spectra is defined as a nonlinear
transformation $\bg_n$ of a linear mixture of $\nbmat$ spectra
$\Vmat{r}$ contaminated by additive noise
\begin{equation}
\label{eq:model0}
    \Vpix{n}  = \bg_n \left( \sum_{\nomat=1}^{\nbmat} {\abond{r}{n} \Vmat{r}
    } \right)+ \Vnoise_n =  \bg_n \left(\MATmat \Vabond{n}\right) + \Vnoise_n
\end{equation}
where $\Vmat{\nomat} =
[\mat{\nomat}{1},\ldots,\mat{\nomat}{\nbband}]\transp$ is the
spectrum of the $\nomat$th material present in the scene,
$\abond{r}{n}$ is its corresponding proportion in the $n$th pixel,
$\nbmat$ is the number of endmembers contained in the image and
$\bg_n$ is a nonlinear function associated with the $n$th pixel.
Moreover, $\Vnoise_n$ is an additive independently distributed
zero-mean Gaussian noise sequence with diagonal covariance matrix
$\bSigma=\diag{\bsigma^2}$, denoted as $\Vnoise_n \sim
\Ndistr{\Vzero_{\nbband},\bSigma}$, where
$\bsigma^2=[\sigma_1^2,\ldots,\sigma_L^2]\transp$ is the vector of
the $L$  noise variances and $\diag{\bsigma^2}$ is an $L \times L$
diagonal matrix containing the elements of the vector $\bsigma^2$.
Note that the usual matrix and vector notations $\MATmat
=[\Vmat{1},\ldots,\Vmat{\nbmat}]$ and
$\Vabond{n}=[\abond{1}{n},\ldots, \abond{\nbmat}{n}]\transp$ have
been used in the right hand side of \eqref{eq:model0}. As in
\cite{Altmann2012}, the $N$ nonlinear functions $\bg_n$ are defined
as second order polynomial nonlinearities defined by
\begin{eqnarray}
\label{eq:gi}
  \bg_n: & [0,1]^L &\rightarrow \mathbb{R}^L\nonumber\\
     & \bfs    &\mapsto  \left[s_1 + \nonlin{n} s_1^2,\ldots,s_L + \nonlin{n} s_L^2\right]\transp
\end{eqnarray}
with $ \mathbf{s} = [s_1,\ldots,s_{\nbband}]\transp$ and
$\nonlin{n}$ is a real parameter. An interesting property of the
resulting nonlinear model referred to as polynomial post nonlinear
mixing model (PPNMM) is that it reduces to the classical LMM for
$\nonlin{n}=0$. Motivations for considering polynomial
nonlinearities have been discussed in \cite{Altmann2012}. In
particular, it has been shown that the PPNMM is very flexible to
approximate many different nonlinearities and can be used for
nonlinearity detection. Straightforward computations allow the PPNMM
observation matrix to be expressed as follows
\begin{equation} \label{eq:model}
\MATpix  =  \MATmat \MATabond + \left[\left( \MATmat \MATabond
\right)\odot \left( \MATmat \MATabond \right)\right] \diag{\Vnonlin}
+ \MATnoise
\end{equation}
where $\MATabond = [\Vabond{1},\ldots,\Vabond{N}]$ is an $R \times
N$ matrix, $\MATpix=[\Vpix{1},\ldots,\Vpix{N}]$ and
$\MATnoise=[\Vnoise_1,\ldots,\Vnoise_N]$ are $L \times N$ matrices,
$\Vnonlin=[\nonlin{1},\ldots,\nonlin{N}]\transp$ is an $N \times 1$
vector containing the nonlinearity parameters and $\odot$ denotes
the Hadamard (termwise) product.

\subsection{Abundance reparametrization}
Due to physical considerations, the abundance vectors $\Vabond{n}$
satisfy the following positivity and sum-to-one constraints
\begin{equation}
\label{eq:abundancesconst}
\sum_{\nomat=1}^{\nbmat}{\abond{r}{n}}=1,~~ \abond{r}{n} > 0,
\forall \nomat \in \left\lbrace 1,\ldots,\nbmat \right\rbrace.
\end{equation}
To handle these constraints, we propose to reparameterize the
abundance vectors belonging to the following set
\begin{equation}
\label{eq:simplex} \Simplex = \left\lbrace
\Vabonds=[\abonds{1},\ldots,\abonds{R}]\transp\left|
\abonds{\nomat}> 0, \sum_{\nomat=1}^{\nbmat}{\abonds{\nomat}} = 1
\right\rbrace \right.
\end{equation}
using the following transformation
\begin{eqnarray}
\abond{r}{n} = \left(\prod_{k=1}^{\nomat-1}{\tabond{k}{n}} \right)
\times \left\{
    \begin{array}{ll}
        1-\tabond{r}{n} & \mbox{if } \nomat < \nbmat \\
        1 & \mbox{if } \nomat = \nbmat
    \end{array}
\right. .
\end{eqnarray}
This transformation has been recently suggested in
\cite{Betancourt2010}. One motivation for using the latent variables
$\tabond{r}{n}$ instead of $\abond{\nomat}{n}$ is the fact that the
constraints \eqref{eq:abundancesconst} for the $n$th abundance
vector $\Vabond{n}$ express as
\begin{eqnarray}
\label{eq:tabundancesconst} 0 < \tabond{r}{n}<1 , \quad \forall r
\in \left \lbrace 1, \ldots, \nbmat-1 \right \rbrace
\end{eqnarray}
for the $n$th coefficient vector
$\Vtabond{n}=[\tabond{1}{n},\ldots,\tabond{\nbmat-1}{n}]\transp$. As
a consequence, the constraints \eqref{eq:tabundancesconst} are much
easier to handle for the sampling procedure than
\eqref{eq:abundancesconst} (as will be shown in Sections
\ref{sec:CHMC} and \ref{sec:Gibbs}). The next section presents the
Bayesian model associated with the PPNMM \eqref{eq:model0} for SU.

\section{Bayesian model}
\label{sec:bayesian} This section generalizes the hierarchical
Bayesian model introduced in \cite{Altmann2012} in order to jointly
estimate the abundances and endmembers, leading to a fully
unsupervised hyperspectral unmixing algorithm. The unknown parameter
vector associated with the PPNMM contains the reparameterized
abundances $\MATtabond=[\Vtabond{1},\ldots,\Vtabond{N}]$ (satisfying
the constraints \eqref{eq:tabundancesconst}), the endmember matrix
$\MATmat$, the nonlinearity parameter vector $\Vnonlin$ and the
additive noise variance $\bsigma^2$. This section summarizes the
likelihood and the parameter priors (associated with the proposed
hierarchical Bayesian PPNMM) introduced to perform nonlinear
unsupervised hyperspectral unmixing.

\subsection{Likelihood}
Equation \eqref{eq:model} shows that $\Vpix{n}|\MATmat,\Vtabond{n},
\nonlin{n}, \bsigma^2$ is distributed according to a Gaussian
distribution with mean $\bg_n\left(\MATmat\Vabond{n}\right)$ and
covariance matrix $\bSigma$, denoted as
$\Vpix{n}|\MATmat,\Vtabond{n}, \nonlin{n}, \bsigma^2 \sim
\Ndistr{\bg_n\left(\MATmat\Vabond{n}\right),\bSigma}$. Note that the
abundance vector $\Vabond{n}$ should be denoted as
$\Vabond{n}(\Vtabond{n})$. However, the argument $\Vtabond{n}$ has
been omitted for brevity. Assuming independence between the observed
pixels, the joint likelihood of the observation matrix $\MATpix$ can
be expressed as
\begin{equation}
\label{eq:likelihood}
    f(\MATpix|\MATmat,\MATtabond, \Vnonlin, \bsigma^2) \propto |\bSigma|^{-N/2}\etr\left[-\dfrac{(\MATpix-\bX)\transp\bSigma^{-1}(\MATpix-\bX)}{2}\right]
\end{equation}
where $\propto$ means ``proportional to'', $\etr(\cdot)$ denotes the
exponential trace and $\bX=\MATmat \MATabond + \left[\left( \MATmat
\MATabond \right)\odot \left( \MATmat \MATabond \right)\right]
\diag{\Vnonlin}$ is an $L \times N$ matrix.

\subsection{Parameter priors}
\subsubsection{Coefficient matrix $\MATtabond$} To reflect the lack of prior
knowledge about the abundances, we propose to assign prior
distributions for the coefficient vector $\Vtabond{n}$ that
correspond to noninformative prior distributions for $\Vabond{n}$.
More precisely, assigning the following beta priors
\begin{eqnarray}
\label{eq:prior_coeff} \tabond{n}{r} \sim \mathcal{B}e(\nbmat-r,1)
\quad r \in \left \lbrace 1, \ldots, \nbmat-1 \right \rbrace
\end{eqnarray}
and assuming prior independence between the elements of
$\Vtabond{n}$ yield an abundance vector $\Vabond{n}$ uniformly
distributed in the set defined in \eqref{eq:simplex} (see
\cite{Betancourt2010} for details). Assuming prior independence
between the coefficient vectors $\left \lbrace \Vtabond{n} \right
\rbrace_{n=1,\ldots,N}$ leads to
\begin{eqnarray}
\label{eq:joint_prior_coeff} f( \MATtabond)  =
\prod_{r=1}^{R-1}\left \lbrace
\dfrac{1}{B(\nbmat-r,1)^{N}}\prod_{n=1}^{N}\tabond{n}{r}^{\nbmat-r-1}\right
\rbrace
\end{eqnarray}
where $B(\cdot,\cdot)$ is the Beta function.
\subsubsection{Endmembers}
Each endmember $\Vmat{r}=[\mat{r}{1},\ldots,\mat{r}{L}]\transp$ is a
reflectance vector satisfying the following constraints
\begin{equation}
\label{eq:endmembersconst}
 0 \leq \mat{r}{\ell} \leq 1, \forall r \in \left\lbrace 1,\ldots,R \right\rbrace , \forall \ell \in \left\lbrace 1,\ldots,L \right\rbrace.
\end{equation}
For each endmember $\Vmat{r}$, we propose to use a Gaussian prior
\begin{eqnarray}
\label{eq:prior_M} \Vmat{r} \sim
\mathcal{N}_{[0,1]^L}(\bar{\bm}_r,s^2\Id{\nbband}),
\end{eqnarray}
 truncated on $[0,1]^L$ to satisfy the constraints \eqref{eq:endmembersconst}. In
this paper, we propose to select the mean vectors $\bar{\bm}_r$ as
the pure components previously identified by the nonlinear EEA
studied in \cite{Heylen2011} and referred to as ``Heylen''. The
variance $s^2$ reflects the degree of confidence given to this prior
information. When no additional knowledge is available, this
variance is fixed to a large value ($s^2=50$ in our simulations).
Note that any EEA could be used to define the vectors
$\bar{\bm}_1,\ldots,\bar{\bm}_R$.
\subsubsection{Nonlinearity parameters}
The PPNMM reduces to the LMM for $\nonlin{n}=0$. Since the LMM is
relevant for most observed pixels, it makes sense to assign prior
distributions to the nonlinearity parameters that enforce sparsity
for the vector $\Vnonlin$. To detect linear and nonlinear mixtures
of the pure spectral signatures in the image, the following
conjugate Bernoulli-Gaussian prior is assigned to the nonlinearity
parameter $\nonlin{n}$
\begin{eqnarray}
f(\nonlin{n} | w, \sigma_b^2)  = (1-w)\delta(\nonlin{n}) + w
\dfrac{1}{\sqrt{2\pi \sigma_b^2}}\exp\left(-\frac{\nonlin{n}^2}{2
\sigma_b^2} \right)
\end{eqnarray}
where $\delta(\cdot)$ denotes the Dirac delta function. Note that
the prior distributions for the nonlinearity parameters $\left
\lbrace \nonlin{n} \right \rbrace_{n=1,\ldots,N}$ share the same
hyperparameters $w \in [0,1]$ and $\sigma_b^2 \in \R^+$. More
precisely, the weight $w$ is the prior probability of having a
nonlinearly mixed pixel in the image. Assuming prior independence
between the nonlinearity parameters $\left \lbrace \nonlin{n} \right
\rbrace_{n=1,\ldots,N}$ , the joint prior distribution of the
nonlinearity parameter vector $\Vnonlin$ can be expressed as follows
\begin{eqnarray}
f(\Vnonlin | w, \sigma_b^2) & = & \prod_{n=1}^{N} f(\nonlin{n} | w,
\sigma_b^2)
\end{eqnarray}
\subsubsection{Noise variances}
A Jeffreys' prior is chosen for the noise variance of each spectral
band $\sigma_{\ell}^2$
\begin{eqnarray}
    \label{eq:priorvar}
    f(\sigma_{\ell}^2) \propto \dfrac{1}{\sigma_{\ell}^2} \Indicfun{\R^+}{\sigma_{\ell}^2}
\end{eqnarray}
which reflects the absence of knowledge for this parameter (see
\cite{Bernardo94} for motivations). Assuming prior independence
between the noise variances, we obtain
\begin{eqnarray}
    \label{eq:jointpriorvar}
    f(\bsigma^2) = \prod_{\ell=1}^L f(\sigma_{\ell}^2).
\end{eqnarray}

\subsection{Hyperparameter priors}
The performance of the proposed Bayesian model for spectral unmixing
depends on the values of the hyperparameters $\sigma_b^2$ and $w$.
When the hyperparameters are difficult to adjust, it is classical to
include them in the unknown parameter vector, resulting in a
hierarchical Bayesian model \cite{Dobigeon2009,Altmann2012}. This
strategy requires to define prior distributions for the
hyperparameters.

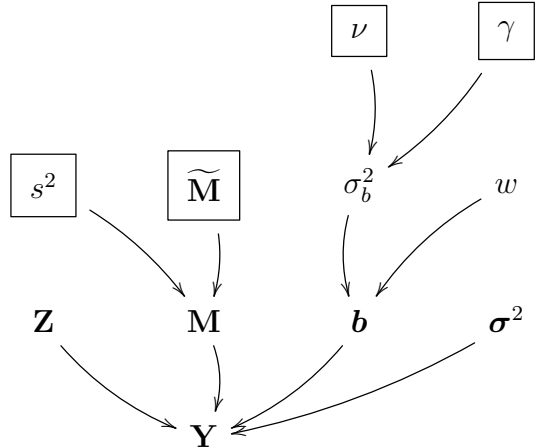
\begin{figure}[h!]
\centerline{ \xymatrix{
  & &*+<0.05in>+[F-]+{\nu} \ar@/^/[d] & *+<0.05in>+[F-]+{\gamma} \ar@/^/[ld]  \\
  *+<0.05in>+[F-]+{s^2} \ar@/^/[rd] & *+<0.05in>+[F-]+{\widetilde{\MATmat}} \ar@/^/[d] & \sigma_b^2 \ar@/_/[d] & w \ar@/_/[ld] \\
  \MATtabond \ar@/_/[rd]    &    \MATmat \ar@/^/[d] &  \Vnonlin \ar@/^/[ld]& \bsigma^2 \ar@/^/[lld]  \\
  & \MATpix &   & }
  }
\caption{DAG for the parameter and hyperparameter priors (the fixed
parameters appear in boxes).} \label{fig:DAG}
\end{figure}

A conjugate inverse-Gamma prior is assigned to $\sigma_b^2$
\begin{eqnarray}
    \label{eq:priorhyper}
    \sigma_b^2 \sim \mathcal{I} \mathcal{G} \left(\gamma, \nu\right)
\end{eqnarray}
where $(\gamma, \nu)$ are real parameters fixed to obtain a flat
prior, reflecting the absence of knowledge about the variance
$\sigma_b^2$ ($(\gamma, \nu)$ will be set to $(10^{-1},10^{-1})$ in
the simulation section). A uniform prior distribution is assigned to
the hyperparameter $w$
\begin{eqnarray}
w \sim \mathcal{U}_{[0,1]}(w)
\end{eqnarray}
since there is no a priori information regarding the proportions of
linearly and nonlinearly mixed pixels in the image. The resulting
directed acyclic graph (DAG) associated with the proposed Bayesian
model is depicted in Fig. \ref{fig:DAG}.

\subsection{Joint posterior distribution}
The  joint posterior distribution of the unknown
parameter/hyperparameter vector $\left\lbrace\paramvect,
\bPhi\right\rbrace$ where $\paramvect = \left\lbrace \MATtabond,
\MATmat, \Vnonlin, \bsigma^2\right\rbrace$ and
$\bPhi=\left\lbrace\sigma_b^2 ,w \right\rbrace$ can be computed
using the following hierarchical structure
\begin{equation}
\label{eq:posterior}
    f(\paramvect,
\bPhi|\MATpix) \propto f(\MATpix|\paramvect, \bPhi) f(\paramvect,
\bPhi)
\end{equation}
where $f(\MATpix|\paramvect)$ has been defined in
\eqref{eq:likelihood}. By assuming \textit{a priori} independence
between the parameters $\MATtabond$, $\MATmat$, $\Vnonlin$ and
$\bsigma^2$ and between the hyperparameters $\sigma_b$ and $w$, the
joint prior distribution of the unknown parameter vector can be
expressed as
\begin{eqnarray}
\label{eq:joint_prior} f(\paramvect, \bPhi) & = & f(\paramvect|
\bPhi)f(\bPhi)\nonumber\\
& = & f(\MATtabond) f(\MATmat) f(\bsigma^2) f(\Vnonlin|\sigma_b^2,w)
f(\sigma_b^2) f(w).
\end{eqnarray}
The joint posterior distribution $f(\paramvect,\bPhi|\MATpix)$ can
then be computed up to a multiplicative constant after replacing
\eqref{eq:joint_prior} and \eqref{eq:likelihood} in
\eqref{eq:posterior}. Unfortunately, it is difficult to obtain
closed form expressions for the standard Bayesian estimators
(including the maximum a posteriori (MAP) and the minimum mean
square error (MMSE) estimators) associated with
\eqref{eq:posterior}. In this paper, we propose to use efficient
Markov Chain Monte Carlo (MCMC) methods to generate samples
asymptotically distributed according to \eqref{eq:posterior}. Due to
the large number of parameters to be sampled, we use an HMC
algorithm which allows the number of sampling steps to be reduced
and which improves the mixing properties of the sampler. The
generated samples are then used to compute the MMSE estimator of the
unknown parameter vector $(\paramvect, \bPhi)$. The next section
summarizes the basic principles of the HMC methods that will be used
to sample asymptotically from \eqref{eq:posterior}.

\section{Constrained Hamiltonian Monte Carlo method}
\label{sec:CHMC} HMCs are powerful methods for sampling from many
continuous distributions by introducing fictitious momentum
variables. Let $\bq \in \mathbb{R}^D$ be the parameter of interest
and $\pi(\bq)$ its corresponding distribution to be sampled from.
From statistical mechanics, the distribution $\pi(\bq)$ can be
related to a potential energy function $U(\bq)=-\log \left[ \pi(\bq)
\right]+ c$ where $c$ is a positive constant such that $\int
\exp\left(-U(\bq)+c \right) \textrm{d}\bq=1$. The Hamiltonian of
$\pi(\bq)$ is a function of the energy $U(\bq)$ and of an additional
momentum vector $\bp \in \mathbb{R}^D$ defined as
\begin{eqnarray}
\label{eq:Hamiltonian} H(\bq,\bp)= U(\bq) + K(\bp)
\end{eqnarray}
where $K(\bp)$ is an arbitrary kinetic energy function. Usually, a
quadratic kinetic energy is chosen and we propose to use
$K(\bp)=\bp\transp\bp/2$ in this paper (for reasons explained
later). The Hamiltonian \eqref{eq:Hamiltonian} defines the following
distribution
\begin{eqnarray}
\label{eq:joint_distr_H}
f(\bq,\bp) & \propto & \exp \left[- H(\bq,\bp)\right]\nonumber \\
          & \propto & \pi(\bq) \exp \left(- \dfrac{1}{2}\bp\transp\bp\right)
\end{eqnarray}
for $(\bq,\bp)$ which shows that $\bq$ and $\bp$ are independent and
that the marginal distribution of $\bp$ is a
$\mathcal{N}(\Vzero_{D},\Id{D})$ distribution. The HMC algorithm
allows samples to be asymptotically generated according to
\eqref{eq:joint_distr_H}. The $i$th HMC iteration starts with an
initial pair of vectors $(\bq^{(i)},\bp^{(i)})$ and consists of two
steps. The first step resamples the initial momentum
$\tilde{\bp}^{(i)}$ according to the standard multivariate Gaussian
distribution. The second step uses Hamiltonian dynamics to propose a
candidate $(\bq^*,\bp^*)$ which is accepted with the following
probability
\begin{eqnarray}
\label{eq:accept_q} \rho = \min \left \lbrace \exp
\left[-H(\bq^*,\bp^*)+H(\bq^{(i)},\tilde{\bp}^{(i)})\right],1 \right
\rbrace.
\end{eqnarray}
\subsection{Generation of the candidate $(\bq^*,\bp^*)$}
\label{subsec:leapfrog} Hamiltonian dynamics are usually simulated
by discretization methods such as Euler or leapfrog methods. The
classical leapfrog method is a discretization scheme composed of
$N_{\textrm{LF}}$ steps with a discretization stepsize $\epsilon$.
The $n$th leapfrog step can be expressed as
\begin{subequations} \label{eq:leapfrog}
\begin{align}
\bp^{(i,n\epsilon+\epsilon/2)} & =  \bp^{(i,n\epsilon)}
                -\dfrac{\epsilon}{2} \dfrac{\partial U}{\partial
                \bq\transp} \left( \bq^{(i,n\epsilon)} \right) \label{eq:leapfrog1} \\
\bq^{(i,(n+1)\epsilon)} & =  \bq^{(i,n\epsilon)} +
                \epsilon \bp^{(i,n\epsilon+\epsilon/2)} \label{eq:leapfrog2} \\
\bp^{(i,(n+1)\epsilon)} & =  \bp^{(i,n\epsilon+\epsilon/2)}
                -\dfrac{\epsilon}{2} \dfrac{\partial
                U}{\partial \bq\transp} \left[ \bq^{(i,(n+1)\epsilon)}
                \right]. \label{eq:leapfrog3}
                \end{align}
\end{subequations}
 The leapfrog method starts with
$(\bq^{(i,0)},\tilde{\bp}^{(i)})=(\bq^{(i)},\tilde{\bp}^{(i)})$ and
the candidate is set after $N_{\textrm{LF}}$ steps to
$(\bq^*,\bp^*)=(\bq^{(i,\epsilon
N_{\textrm{LF}})},\tilde{\bp}^{(i,\epsilon N_{\textrm{LF}})})$.

However, if $\bq$ is subject to constraints, more sophisticated
discretization methods must be used. Assume that the vector of
interest $\bq=[q_1,\ldots,q_D]\transp$ satisfies the following
constraints
\begin{eqnarray}
\label{eq:constraints_q} q_l<q_d<q_u,\quad d\in \left \lbrace
1,\ldots,D\right \rbrace
\end{eqnarray}
where $q_l$ (resp. $q_u$) is the lower (resp. upper) bound for $q_d$
(such kind of constraints need to be satisfied by the elements of
$\MATtabond$ and the endmembers in $\MATmat$). In this paper we
propose to use the constrained leapfrog scheme studied in
\cite[Chap. 5]{Brooks2011}, consisting of $N_{\textrm{LF}}$ steps,
with a discretization stepsize $\epsilon_q$. Each CHMC iteration
starts in a similar way to the classical leapfrog method, with the
sequential sampling of the momentum $\bp$ \eqref{eq:leapfrog1} and
the vector $\bq$ \eqref{eq:leapfrog2}. However, if the generated
vector $\bq$ violates the constraints \eqref{eq:constraints_q}, it
is modified depending on the violated constraints and the momentum
is negated (see \cite[Chap. 5]{Brooks2011} for more details). This
step is repeated until each component of the generated $\bq$
satisfies the contraints. The CHMC ends with the update of the
momentum $\bp$ \eqref{eq:leapfrog3}. One iteration of the resulting
constrained HMC algorithm (CHMC) is summarized in Algo.
\ref{algo:HMC_q}. As mentioned above, one might think of using a
more sophisticated kinetic energy for $\bp$ to improve the
performance of the HMC algorithm. However, the kinetic energy
$K(\bp)=\bp\transp\bp/2$ allows the discretization method handling
the constraints to be simple and will provide good performance for
our application (as will be shown in Section \ref{sec:simu_synth}).
The performance of the HMC mainly relies on the values of the
parameters $N_{\textrm{LF}}$ and $\epsilon_q$. Fortunately, the
choice of $\epsilon_q$ is almost independent of $N_{\textrm{LF}}$
such that these two parameters can be tuned sequentially. The
procedures used in this paper to adjust $N_{\textrm{LF}}$ and
$\epsilon_q$ are detailed in the next paragraphs.
\subsection{Tuning the stepsize $\epsilon_q$}
\label{subsec:tuning_stepsize} The step size $\epsilon_q$ is related
to the accuracy of the leapfrog method to approximate the
Hamiltonian dynamics. When $\epsilon_q$ is ``small'', the
approximation of the Hamiltonian dynamic is accurate and the
acceptance rate \eqref{eq:accept_q} is high. However, the
exploration of the distribution support is slow (for a given
$N_{\textrm{LF}}$). In this paper, we propose to tune the stepsize
during the burn-in period of the sampler. More precisely, the
stepsize is decreased (resp. increased) by $25\%$ if the average
acceptance rate over the last $50$ iterations is smaller than $0.5$
(resp. higher than $0.8$). Note that the stepsize update only
happens during the burn-in period to ensure the Markov chain is
homogeneous after the burn-in period.

\subsection{Tuning the number of leapfrog steps $N_{\textrm{LF}}$}
\label{subsec:tuning_nbleapfrog} Assume $\epsilon_q$ has been
correctly adjusted. Too small values of $N_{\textrm{LF}}$ lead to a
slow exploration of the distribution (random walk behavior) whereas
too high values of $N_{\textrm{LF}}$ require high computational
time. Similarly to the stepsize $\epsilon_q$, the optimal choice of
$N_{\textrm{LF}}$ depends on the distribution to be sampled. The
sampling procedure proposed in this paper consists of several HMC
updates included in a Gibbs sampler (as will be shown in the next
section). The number of leapfrog steps required for each of these
CHMC updates has been adjusted by cross-validation. From preliminary
runs, we have observed that setting the number of leapfrog steps for
each HMC update close to $N_{\textrm{LF}}=50$ provides a reasonable
tradeoff ensuring a good exploration of the target distribution and
a reasonable computational complexity. To avoid possible periodic
trajectories, it is recommended to let $N_{\textrm{LF}}$ random
\cite[Chap. 5]{Brooks2011}. In this paper, we have assumed that
$N_{\textrm{LF}}$ is uniformly drawn in the interval $[45,55]$ at
each iteration of the Gibbs sampler. The next section presents the
Gibbs sampler (including CHMC steps) which is proposed to sample
according to \eqref{eq:posterior}.

\begin{figure}[h!]
\begin{algogo}{Constrained
Hamiltonian Monte Carlo iteration} \small
     \label{algo:HMC_q}
     \begin{algorithmic}[1]
        \STATE \underline{\%Initialization of the $i$th iteration($n=0$)}
        \begin{itemize}
        \item $\bq^{(i,0)}=\bq^{(i)}$ satisfying the constraints
            \eqref{eq:constraints_q}
        \item Sample $\bp^{(i,0)}=\tilde{\bp}^{(i)}) \sim
            \Ndistr{\Vzero_{D}, \Id{D}}$
        \end{itemize}
        \STATE \underline{\%Modified leapfrog steps}
        \FOR{$n=0:N_{\textrm{LF}}-1$}
        \STATE \underline{\%Standard leapfrog steps}
        \STATE
        \begin{itemize}
            \item Compute $\bp^{(i,n\epsilon+\epsilon/2)} =
                \bp^{(i,n\epsilon)} -\dfrac{\epsilon}{2}
                \dfrac{\partial U}{\partial \bq\transp} \left(
                \bq^{(i,n\epsilon)} \right)$
            \item Compute $\bq^{(i,(n+1)\epsilon)}  =
                \bq^{(i,n\epsilon)} + \epsilon
                \bp^{(i,n\epsilon+\epsilon/2)}$
            \end{itemize}
        \STATE \underline{\%Steps required to ensure $\bq^{(i,(n+1)\epsilon)}$ satisfies \eqref{eq:constraints_q}}
        \WHILE{$\bq^{(i,(n+1)\epsilon)}$ does not satisfy \eqref{eq:constraints_q}}
        \FOR{$d=1:D$}
        \IF{$q_d^{(i,(n+1)\epsilon)}<q_l$}
        \STATE Set  $q_d^{(i,(n+1)\epsilon)} = 2q_l -
        q_d^{(i,(n+1)\epsilon)}$\\
        (replace $q_d^{(i,(n+1)\epsilon)}$ by its symmetric with
        respect to $q_l$)
        \STATE Set  $p_d^{(i,n\epsilon+\epsilon/2)} = - p_d^{(i,n\epsilon+\epsilon/2)}$
        \ENDIF
        \IF{$q_d^{(t+\epsilon)}>q_u$}
        \STATE Set  $q_d^{(i,(n+1)\epsilon)} = 2q_u - q_d^{(i,(n+1)\epsilon)}$\\
        (replace $q_d^{(i,(n+1)\epsilon)}$ by its symmetric with
        respect to $q_u$)
        \STATE Set  $p_d^{(i,n\epsilon+\epsilon/2)} = - p_d^{(i,n\epsilon+\epsilon/2)}$
        \ENDIF
        \ENDFOR
        \ENDWHILE
        \STATE \underline{\%Standard leapfrog step}
        \STATE Compute $\bp^{(i,(n+1)\epsilon)}  =  \bp^{(i,n\epsilon+\epsilon/2)}
                -\dfrac{\epsilon}{2} \dfrac{\partial
                U}{\partial \bq\transp} \left[ \bq^{(i,(n+1)\epsilon)}
                \right]$
        \ENDFOR
        \STATE \underline{\%Accept-reject procedure}
        \STATE Set $\bp^{*}=\bp^{(i,\epsilon
N_{\textrm{LF}})}$ and $\bq^{*}=\bq^{(i,\epsilon N_{\textrm{LF}})}$
        \STATE Compute $\rho$ using \eqref{eq:accept_q}
        \STATE Set $(\bq^{(i+1)},\bp^{(i+1)})=(\bq^{*},\bp^{*})$ with probability $\rho$
        \STATE Else set
        $(\bq^{(i+1)},\bp^{(i+1)})=(\bq^{(i)},\tilde{\bp}^{(i)})$.
\end{algorithmic}
\end{algogo}
\end{figure}
\clearpage

\section{Gibbs Sampler}
\label{sec:Gibbs} The principle of the Gibbs sampler is to sample
according to the conditional distributions of the posterior of
interest \cite[Chap. 10]{Robert2004}. Due to the large number of
parameters to be estimated, it makes sense to use a block Gibbs
sampler to improve the convergence of the sampling procedure. More
precisely, we propose to sample sequentially $\MATmat, \MATtabond,
\Vnonlin, \bsigma^2, \sigma_b^2$ and $w$ using six moves that are
detailed in the next sections.
\subsection{Sampling the coefficient matrix $\MATtabond$}
 \label{subsec:sample_z}
Sampling from
$f(\MATtabond|\MATpix,\MATmat,\Vnonlin,\bsigma^2,\sigma_b^2,w)$ is
difficult due to the complexity of this distribution. In this case,
it is classical to use an accept/reject procedure to update the
coefficient matrix $\MATtabond$ (leading to a hybrid
Metropolis-Within-Gibbs sampler). Since the elements of $\MATtabond$
satisfy the constraints \eqref{eq:tabundancesconst}, the CHMC
studied in Section \ref{sec:CHMC} could be used to sample according
to the conditional distribution
$f(\MATtabond|\MATpix,\MATmat,\Vnonlin,\bsigma^2,\sigma_b,w)$.
However, as for Metropolis-Hastings updates, the convergence of HMCs
generally slows down when the dimensionality of the vector to be
sampled increases. Consequently, sampling an $N(R-1)$-dimensional
vector using the proposed CHMC can be inefficient when the number of
pixels is very large. However, it can be shown that
\begin{eqnarray}
f(\MATtabond|\MATpix,\MATmat,\Vnonlin,\bsigma^2,\sigma_b,w) =
\prod_{n=1}^{N} f(\Vtabond{n}|\Vpix{n}, \MATmat, \nonlin{n},
\bsigma^2),
\end{eqnarray}
i.e., the $N$ coefficients vectors $\left \lbrace \Vtabond{n} \right
\rbrace_{n=1,\ldots,N}$ are a posteriori independent and can be
sampled independently in a parallel manner. Straightforward
computations lead to
\begin{eqnarray}
    \label{eq:posttabond}
    f(\Vtabond{n}|\Vpix{n}, \MATmat, \nonlin{n},
\bsigma^2) & \propto &
    \exp \left(- \dfrac{(\Vpix{n}-\bx_n)\transp \bSigma^{-1} (\Vpix{n}-\bx_n)}{2}
    \right)\nonumber \\
    & \times & \Indicfun{(0,1)^{\nbmat-1}}{\Vtabond{n}} \prod_{r}^{R-1} \tabond{n}{r}^{\nbmat-r-1}
\end{eqnarray}
where $\bx_n=\bg_n\left(\MATmat\Vabond{n}\right)$,
$\Indicfun{(0,1)^{\nbmat-1}}{\cdot}$ denotes the indicator function
over $(0,1)^{\nbmat-1}$. The distribution \eqref{eq:posttabond} is
related to the following potential energy
\begin{eqnarray}
\label{eq:potential_z}
U(\Vtabond{n})&  = & \dfrac{(\Vpix{n}-\bx_n)\transp \bSigma^{-1} (\Vpix{n}-\bx_n)}{2}\nonumber\\
& - & \sum_{r=1}^{R-1} \log\left(\tabond{n}{r}^{\nbmat-r-1}\right)
\end{eqnarray}
where we note that $f(\Vtabond{n}|\Vpix{n}, \MATmat, \nonlin{n},
\bsigma^2) \propto \exp \left[-U(\Vtabond{n})\right]$. $N$ momentum
vectors associated with a canonical kinetic energy are introduced.
The CHMC of Section \ref{sec:CHMC} is then applied independently to
the $N$ vectors $\Vtabond{n}$ whose dimension ($R-1$) is relatively
small. The partial derivatives of the potential function
\eqref{eq:potential_z} required in Algo. \ref{algo:HMC_q} are
derived in the Appendix.

\subsection{Sampling the endmember matrix $\MATmat$}
From \eqref{eq:posterior} and \eqref{eq:joint_prior}, it can be seen
that
\begin{eqnarray}
f(\MATmat|\MATpix, \MATtabond, \Vnonlin,
\bsigma^2,s^2,\widetilde{\MATmat}) =
\prod_{\ell=1}^{\nbband}{f(\Vmat{\ell,:}|\Vpix{\ell,:},\MATtabond,\Vnonlin,\sigma_{\ell}^2,s^2,\bar{\bm}_{\ell,:})}\nonumber
\end{eqnarray}
where $\Vmat{\ell,:}$ (resp. $\bar{\bm}_{\ell,:}$ and
$\Vpix{\ell,:}$) is the $\ell$th row of $\MATmat$ (resp. of
$\widetilde{\MATmat}$ and $\MATpix$) and
$$f(\Vmat{\ell,:}|\Vpix{\ell,:},\MATtabond,\Vnonlin,\sigma_{\ell}^2,s^2,\bar{\bm}_{\ell,:}) \propto \exp \left(- \dfrac{\Vert \Vpix{\ell,:}- \btt_{\ell} \Vert ^2}{2\sigma_{\ell}^2} \right)$$
\begin{eqnarray}
\label{eq:postendmembers}
  & \times & \exp \left(- \dfrac{\Vert \Vmat{\ell,:}- \bar{\bm}_{\ell,:} \Vert ^2}{2s^2} \right) \Indicfun{(0,1)^{\nbmat}}{\Vmat{\ell,:}}
\end{eqnarray}
with $\btt_{\ell}=\MATabond\transp \Vmat{\ell,:} +
\diag{\Vnonlin}\left[\left(\MATabond\transp \Vmat{\ell,:} \right)
\odot \left(\MATabond\transp \Vmat{\ell,:} \right) \right]$.
Consequently, the rows of the endmember matrix $\MATmat$ can be
sampled independently similarly to the procedure described in the
previous section (to sample $\MATtabond$). More precisely, we
introduce a potential energy $V(\Vmat{\ell,:})$ associated with
$\Vmat{\ell,:}$ defined by
\begin{eqnarray}
\label{eq:potential_m} V(\Vmat{\ell,:})&  = & \dfrac{\Vert
\Vpix{\ell,:}- \btt_{\ell} \Vert ^2}{2\sigma_{\ell}^2} +
\dfrac{\Vert \Vmat{\ell,:}- \bar{\bm}_{\ell,:} \Vert ^2}{2s^2}
\end{eqnarray} and a momentum vector associated with a canonical kinetic energy.
The partial derivatives of the potential function
\eqref{eq:potential_m} required in Algo. \ref{algo:HMC_q} are
derived in the Appendix.

\subsection{Sampling the nonlinearity parameter vector $\Vnonlin$}
\label{subsec:sample_b} Using \eqref{eq:posterior} and
\eqref{eq:joint_prior}, it can be easily shown that the conditional
distribution of $\nonlin{n}|\Vpix{n},\MATmat
\Vtabond{n},\bsigma^2,w,\sigma_b^2$ is the following
Bernoulli-Gaussian distribution
\begin{equation}
    \label{eq:postb}
    \nonlin{n}|\Vpix{n}, \MATmat, \Vtabond{n},\bsigma^2,w,\sigma_b^2 \sim (1-w_n^*)\delta(\nonlin{n}) + w_n^* \mathcal{N}\left(\mu_n,
    s_n^2 \right)
\end{equation}
where
\begin{equation*}
    \mu_n = \dfrac{\sigma_b^2 \left(\Vpix{n} - \MATmat \Vabond{n} \right)\transp\bSigma^{-1}\bh_n}{\sigma_b^2 \bh_n\transp\bSigma^{-1} \bh_n + 1} , \quad
    s_n^2 = \dfrac{\sigma_b^2}{\sigma_b^2 \bh_n\transp\bSigma^{-1} \bh_n + 1}
\end{equation*}
and $\bh_n=(\MATmat\Vabond{n})\odot(\MATmat\Vabond{n})$. Moreover,
\begin{eqnarray}
w_n^* & = & \dfrac{w}{\beta_n + w(1-\beta_n)}\nonumber\\
\beta_n & = & \dfrac{\sigma_b}{s_n} \exp \left(-\dfrac{\mu_n^2}{2
s_n^2} \right).
\end{eqnarray}
For each $\nonlin{n}$, the conditional distribution \eqref{eq:postb}
does not depend on $\left \lbrace \nonlin{k} \right \rbrace_{k \neq
n}$. Consequently, the nonlinearity parameters $\left \lbrace
\nonlin{n} \right \rbrace_{n=1,\ldots,N}$ can be sampled
independently in a parallel manner.

\subsection{Sampling the noise variance vector $\bsigma^2$}
By considering the posterior distribution \eqref{eq:posterior}, it
can be shown that
\begin{eqnarray}
f(\bsigma^2|\MATpix, \MATmat, \MATtabond,\Vnonlin) =
\prod_{\ell=1}^{L} f(\sigma_{\ell}^2|\Vpix{\ell,:},
\Vmat{:,\ell},\MATtabond,\Vnonlin)
\end{eqnarray}
and that $\sigma_{\ell}^2|\Vpix{\ell,:},
\Vmat{:,\ell},\MATtabond,\Vnonlin$ is distributed according to the
following inverse-gamma distribution
\begin{eqnarray}
    \label{eq:postsigma2}
    \sigma_{\ell}^2|\Vpix{\ell,:}, \Vmat{:,\ell},\MATtabond,\Vnonlin \sim \calI \calG \left(\dfrac{N}{2},
    \dfrac{(\Vpix{\ell,:}-\bx_{\ell,:})\transp(\Vpix{\ell,:}-\bx_{\ell,:})}{2}\right)
\end{eqnarray}
where $\bX=[\bx_{1,:},\ldots,\bx_{L,:}]\transp$. Thus the noise
variances can be sampled easily and independently.

\begin{figure}[h!]
\begin{algogo}{Gibbs sampler}
     \label{algo:Gibbs}
     \begin{algorithmic}[1]
        \STATE \underline{Initialization $t=0$}
        \begin{itemize}
        \item $\MATtabond^{(0)},\MATmat^{(0)},\Vnonlin^{(0)},
            \bsigma^{2(0)},w^{(0)},\sigma_b^{2(0)}$.
        \end{itemize}
        \STATE \underline{Iterations}
        \FOR{$t=1:N_{\textrm{MC}}$}
        \STATE \underline{Parameter update}
        \STATE Sample $\MATtabond^{(t)}$ from the pdfs
        \eqref{eq:posttabond} using a CHMC procedure.
        \STATE Sample $\MATmat^{(t)}$ from the pdfs
        \eqref{eq:postendmembers} using a CHMC procedure.
        \STATE Sample $\Vnonlin^{(t)}$ from the pdfs
        \eqref{eq:postb}.
        \STATE Sample $\bsigma^{2(t)}$ from the pdfs
        \eqref{eq:postsigma2}.
        \STATE \underline{Hyperparameter update}
        \STATE Sample $\sigma_b^{2(t)}$ from the pdf
        \eqref{eq:postsigma2b}.
        \STATE Sample $w^{(t)}$ from the pdf
        \eqref{eq:postw}.
        \ENDFOR
\end{algorithmic}
\end{algogo}
\end{figure}

\subsection{Sampling the hyperparameters $\sigma_b^2$ and $w$}
\label{subsec:sample_hyperparameters} Looking carefully at the
posterior distribution \eqref{eq:posterior}, it can be seen that
$\sigma_b^2|\Vnonlin,\gamma, \nu$ is distributed according to the
following inverse-gamma distribution
\begin{eqnarray}
    \label{eq:postsigma2b}
    \sigma_b^2|\Vnonlin,\gamma, \nu \sim \calI \calG \left(\dfrac{n_1}{2}+\gamma,
    \sum_{n \in I_1}\dfrac{\nonlin{n}^2}{2} + \nu \right)
\end{eqnarray}
with $I_1=\left \lbrace n|\nonlin{n}\neq 0 \right \rbrace$,
$n_0=\norm{\Vnonlin}_0$ (where $\norm{\cdot}_0$ is the $\ell_0$
norm, i.e., the number of elements of $\Vnonlin$ that are different
from zero) and $n_1 = N-n_0$, from which it is easy to sample.
Similarly, we obtain
\begin{eqnarray}
\label{eq:postw} w|\Vnonlin \sim \mathcal{B}e(n_1+1,n_0+1).
\end{eqnarray}
Finally, the Gibbs sampler (including HMC procedures) used to sample
according to the posterior \eqref{eq:posterior} consists of the six
steps summarized in Algo. \ref{algo:Gibbs}. The small number of
sampling steps is due to the high parallelization properties of the
proposed sampling procedure, i.e., the generation of the $N$
coefficient vectors $\left \lbrace \Vtabond{n} \right
\rbrace_{n=1,\ldots,N}$, the $N$ nonlinearity parameters $\left
\lbrace \nonlin{n} \right \rbrace_{n=1,\ldots,N}$ and the $\nbband$
reflectance vectors $\left \lbrace \Vmat{\ell,:} \right
\rbrace_{\ell=1,\ldots,\nbband}$. After generating $N_{\textrm{MC}}$
samples using the procedures detailed above, the MMSE estimator of
the unknown parameters can be approximated by computing the
empirical averages of these samples, after an appropriate burn-in
period\footnote{The length of the burn-in period has been determined
using appropriate convergence diagnoses \cite{Robertmcmc}.}. The
next section studies the performance of the proposed algorithm for
synthetic hyperspectral images.

\section{Simulations on synthetic data}
\label{sec:simu_synth} \subsection{Simulation scenario} The
performance of the proposed nonlinear SU algorithm is first
evaluated by unmixing 3 synthetic images of size $50 \times 50$
pixels. The $\nbmat = 3$ endmembers observed at $L=207$ different
spectral bands and contained in these images have been extracted
from the spectral libraries provided with the ENVI software
\cite{ENVImanual2003} (i.e., green grass, olive green paint and
galvanized steel metal). The first synthetic image $I_1$ has been
generated using the standard linear mixing model (LMM). A second
image $I_2$ has been generated according to the PPNMM and a third
image $I_3$ has been generated according to the generalized bilinear
mixing model (GBM) presented in \cite{Halimi2010}. For each image,
the abundance vectors $\Vabond{\nopix}, n=1,\ldots,2500$ have been
randomly generated according to a uniform distribution in the
admissible set defined by
\begin{equation}
\label{eq:truncated_simplex} \Simplex_{\mathrm{t}} = \left\lbrace
\Vabonds\left| 0 <\abonds{\nomat}< 0.9,
\sum_{\nomat=1}^{\nbmat}{\abonds{\nomat}} = 1 \right\rbrace \right..
\end{equation}
Note that the conditions $\abonds{\nomat}< 0.9$ ensure that there is
no pure pixel in the images, which makes the unmixing problem more
challenging. All images have been corrupted by an additive
independent and identically distributed (i.i.d) Gaussian noise of
variance $\noisevar = 10^{-4}$, corresponding to an average
signal-to-noise ratio $\mathrm{SNR} \simeq 21$dB for the three
images. The noise is assumed to be i.i.d. to fairly compare unmixing
performance with SU algorithms assuming i.i.d. Gaussian noise. The
nonlinearity coefficients are uniformly drawn in the set $[0,1]$ for
the GBM. The parameters $\nonlin{n},n=1,\ldots,N$ have been
generated uniformly in the set $[-0.3,0.3]$ for the PPNMM.

\subsection{Comparison with other SU procedures} Different
estimation procedures have been considered for the three mixing
models. More precisely,
\begin{itemize}
\item Two unmixing algorithms have been considered for the LMM.
    The first strategy extracts the endmembers from the whole
    image using the N-FINDR algorithm \cite{Winter1999} and
    estimates the abundances using the FCLS algorithm
    \cite{Heinz2001} (it is referred to as ``SLMM'' for
    supervised LMM). The second strategy is a Bayesian algorithm
    which jointly estimates the endmembers and the abundance
    matrix \cite{Dobigeon2009} (it
     is referred to as ``ULMM'' for unsupervised LMM).
\item Two approaches have also been considered for the PPNMM.
    The first strategy uses the nonlinear EEA studied in
    \cite{Heylen2011} and the gradient-based approach based on
    the PPNMM studied in \cite{Altmann2012} for estimating the
    abundances and the nonlinearity parameter. This strategy is
    referred to as ``SPPNMM'' (supervised PPNMM). The second
    strategy is the proposed unmixing procedure referred to as
    ``UPPNMM'' (unsupervised PPNMM).
\item The unmixing strategy used for the GBM is the nonlinear
    EEA studied in \cite{Heylen2011} and the gradient-based
    algorithm presented in \cite{Halimi2011IGARSS} for abundance
    estimation.
\end{itemize}

The quality of the unmixing procedures can be measured by comparing
the estimated and actual abundance vector using the root normalized
mean square error (RNMSE) defined by
\begin{eqnarray}
\label{eq:RMSE}
    \textrm{RNMSE}= \sqrt{\dfrac{1}{\nbpix\nbmat}\sum_{\nopix=1}^{\nbpix}
    {\norm{\hat{\Vabonds}_{\nopix} - \Vabond{\nopix}}^2}}
\end{eqnarray}
where $\Vabond{\nopix}$ and $\hat{\Vabonds}_{\nopix}$ are the actual
and estimated abundance vectors for the $\nopix$th pixel of the
image and $\nbpix$ is the number of image pixels. Table
\ref{tab:RMSE_synth} shows the RNMSEs associated with the images
$I_1,\ldots,I_3$ for the different estimation procedures. These
results show that the proposed UPPNMM performs better (in term of
RNMSE) than the other considered unmixing methods for the three
images. Moreover, the proposed method provides similar results when
compared with the ULMM for the linearly mixed image $I_1$.

\begin{table}[h!]
\renewcommand{\arraystretch}{1.2}
\begin{footnotesize}
\begin{center}
\caption{Abundance RNMSEs ($\times 10^{-2}$): synthetic
images.\label{tab:RMSE_synth}}
\begin{tabular}{|c|c|c|c|c|}
\cline{3-5}
\multicolumn{2}{c|}{}  &  $I_1$  & $I_2$  &  $I_3$  \\
\multicolumn{2}{c|}{}  &  (LMM)  & (PPNMM)&  (GBM)  \\
\hline
\multirow{2}*{LMM}  & SLMM & $3.78$ & $13.21$ & $6.83$\\
 \cline{2-5}
   &  ULMM & $0.66 $& $10.87$ & $4.21$ \\
\hline
 \multirow{2}*{PPNMM}  & SPPNMM & $4.18$& $6.04$ & $4.13$\\
\cline{2-5}
   &  UPPNMM & $\textbf{\blue{0.37}}$ & $\textbf{\blue{0.81}}$ & $\textbf{\blue{1.38}}$ \\
\hline
\multicolumn{2}{|c|}{GBM} & $4.18$& $11.15$& $5.02$\\
\hline
\end{tabular}
\end{center}
\end{footnotesize}
\vspace{-0.4cm}
\end{table}

\begin{figure}[h!]
\begin{minipage}[b]{1.0\linewidth}
  \centering
\includegraphics[width=0.45\linewidth]{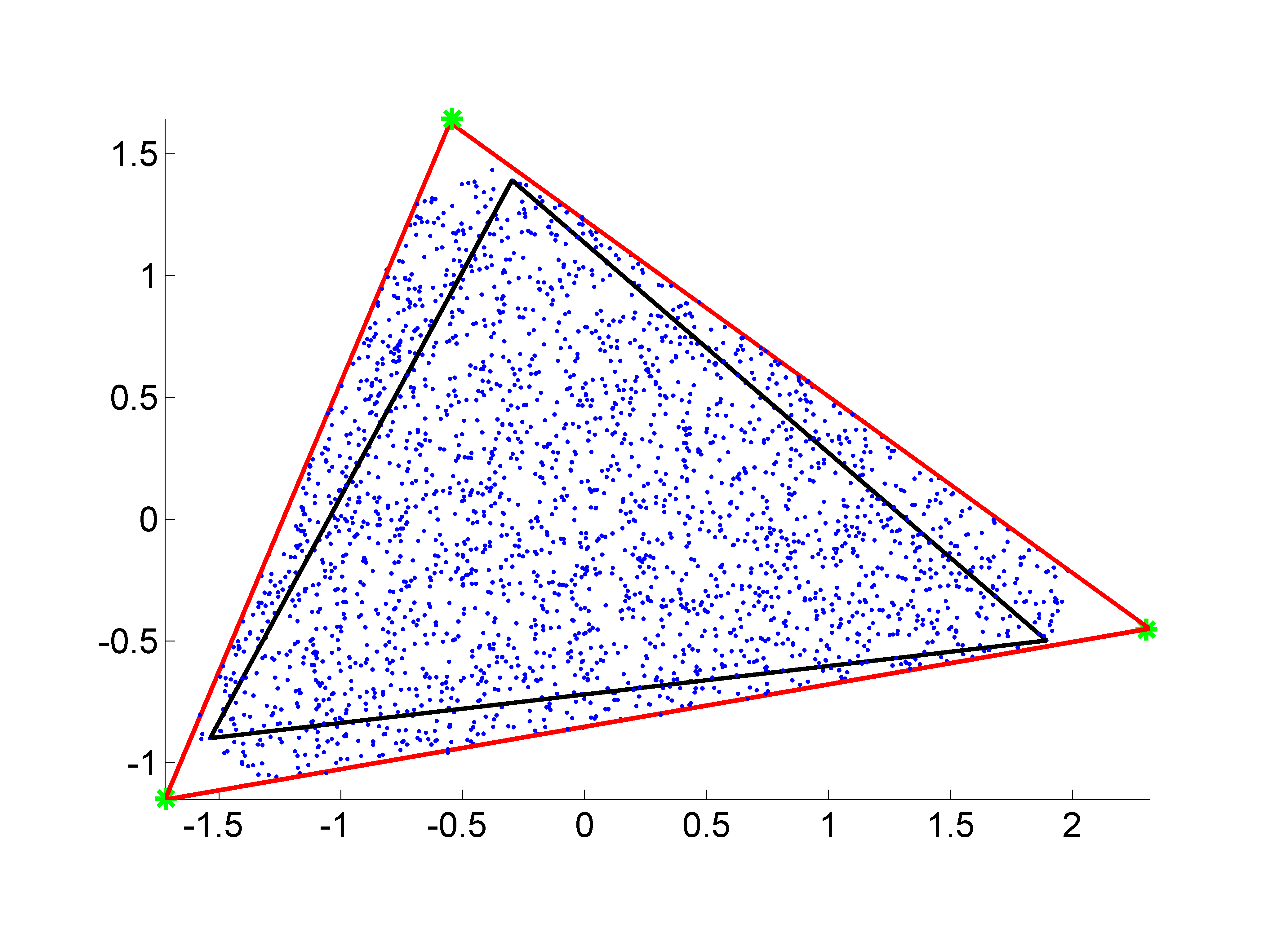}
  \centerline{(a) $I_1$}\medskip
\end{minipage}
\begin{minipage}[b]{0.48\linewidth}
  \centering
\includegraphics[width=\linewidth]{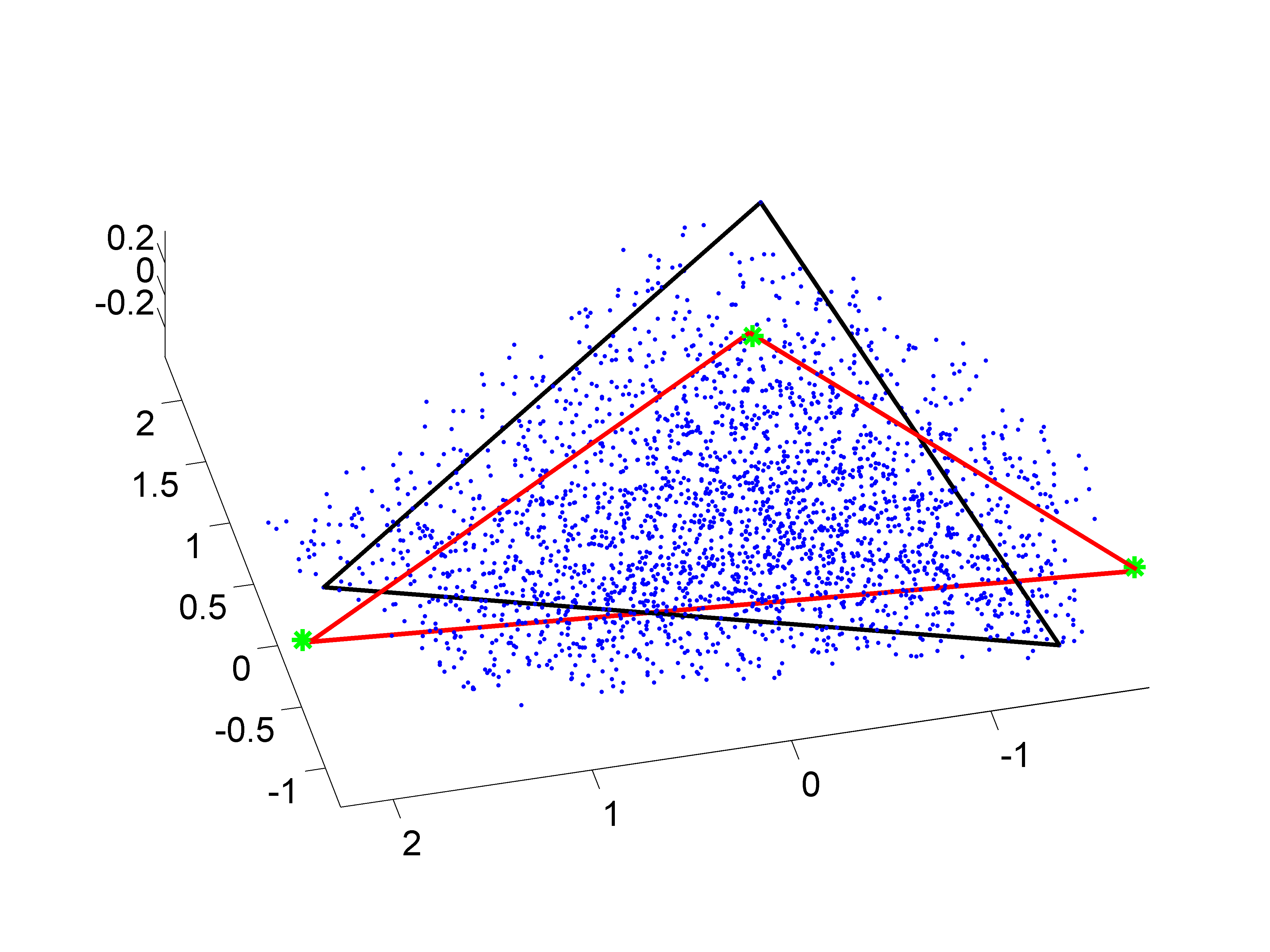}
  \centerline{(b) $I_2$}\medskip
\end{minipage}
\hfill
\begin{minipage}[b]{0.48\linewidth}
  \centering
\includegraphics[width=\linewidth]{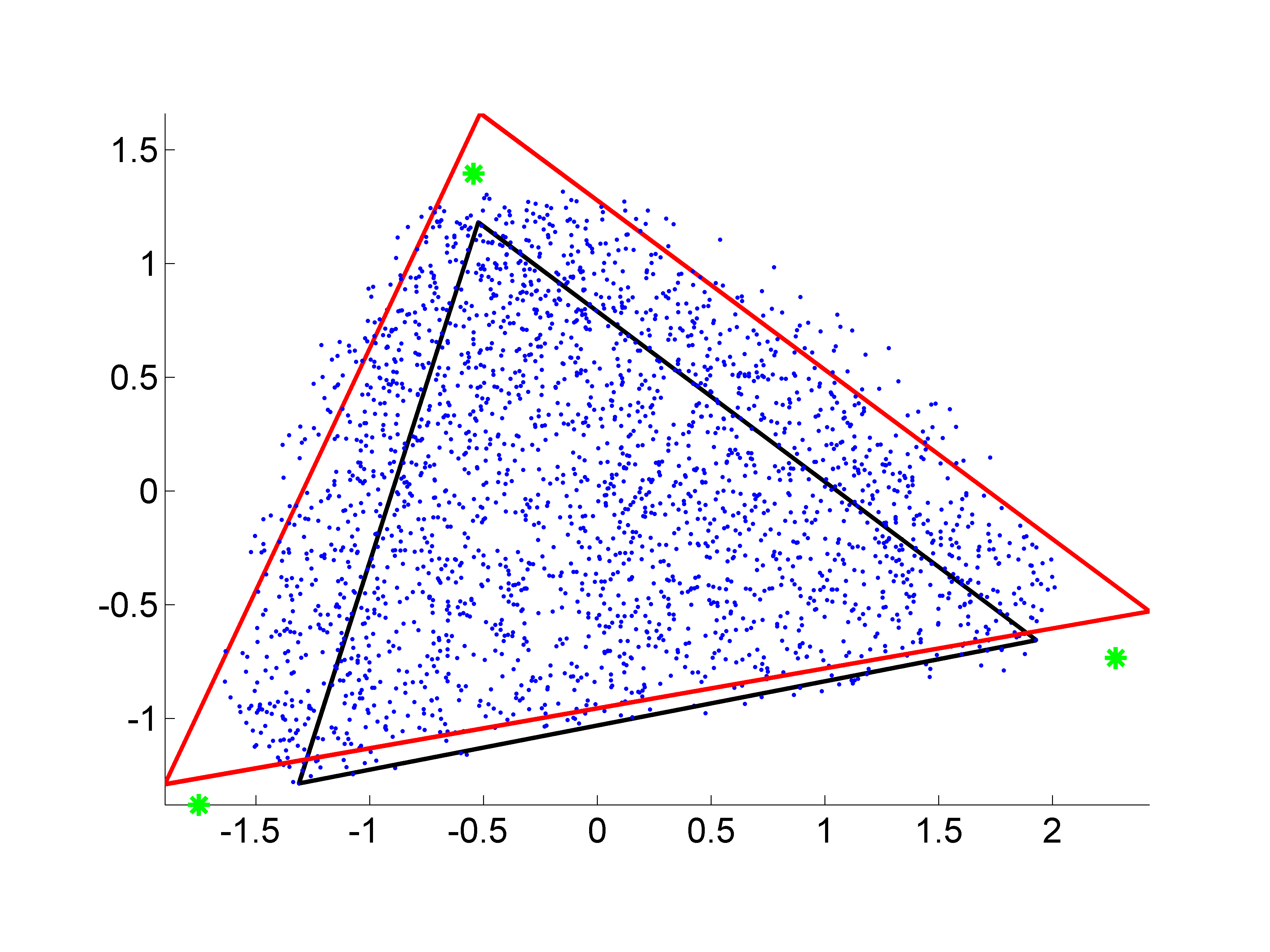}
  \centerline{(c) $I_3$}\medskip
\end{minipage}
\vspace{-0.3cm} \caption{Visualization of the $N=2500$ pixels (blue
dots) of $I_1$, $I_2$ and $I_3$ using the first principal components
  provided by the standard PCA. The green stars correspond to the actual endmembers and the triangles are
  the simplexes defined by the endmembers estimated by the Heylen's method (black) and the proposed method (red).}
\label{fig:simplexes}
\end{figure}

Fig. \ref{fig:simplexes} compares the endmember simplexes estimated
by Heylen's method \cite{Heylen2011} (black) (used to build the
endmember prior) and by the proposed method (red) to the actual
endmembers (green stars). For visualization, the observed pixels and
the actual and estimated endmembers have been projected onto the
three first axes provided by the principal component analysis. These
figures show that the proposed unmixing procedure provides accurate
estimated endmembers for the three images $I_1$ to $I_3$. Due to the
absence of pure pixels in the image, the manifold generated by the
observed pixels $\MATpix$ is difficult to estimate. This explains
the limited performance obtained with Heylen's method. Conversely,
the use of the prior \eqref{eq:prior_M} allows the endmembers
$\Vmat{r}$ to depart from the prior estimations $\bar{\bm}_r$
leading to improved performance.

The quality of endmember estimation is also evaluated by the
spectral angle mapper (SAM) defined as
\begin{eqnarray}
\label{eq:SAM}
    \textrm{SAM}= \textrm{arccos}\left(\dfrac{\left \langle \hat{\bm}_r, \Vmat{r}\right \rangle}{\norm{\hat{\bm}_r} \norm{\Vmat{r}}} \right)
\end{eqnarray}
where $\Vmat{r}$ is the $r$th actual endmember and $\hat{\bm}_r$ its
estimate. The smaller $|\textrm{SAM}|$, the closer the estimated
endmembers to their actual values. Table \ref{tab:SAM_synth}
compares the performance of the different endmember estimation
algorithms. This table shows that the proposed UPPNMM generally
provides more accurate endmember estimates than the others methods.
Moreover, these results illustrate the robustness of the PPNMM
regarding model mis-specification. Note that the ULMM and the UPPNMM
provide similar results (in term of SAMs) for the image $I_1$
generated according to the LMM.

Finally, the unmixing quality can be evaluated by the reconstruction
error (RE) defined as
\begin{eqnarray}
\label{eq:RE}
    \textrm{RE}= \sqrt{\dfrac{1}{\nbpix \nbband}\sum_{\nopix=1}^{\nbpix}
    {\norm{\hat{\Vpixels}_{\nopix} - \Vpix{\nopix}}^2}}
\end{eqnarray}
where $\Vpix{\nopix}$ is the $\nopix$th observation vector and
$\hat{\Vpixels}_{\nopix}$ its estimate. Table \ref{tab:RE_synth}
compares the REs obtained for the different synthetic images. These
results show that the REs are close for the different unmixing
algorithms even if the estimated abundances can vary more
significantly (see Table \ref{tab:RMSE_synth}). Again, the proposed
PPNMM seems to be more robust than the other mixing models to
deviations from the actual model in term of RE.

\begin{table}[h!]
\renewcommand{\arraystretch}{1.2}
\begin{footnotesize}
\begin{center}
\caption{SAMs ($\times 10^{-2}$): synthetic
images.\label{tab:SAM_synth}}
\begin{tabular}{|c|c|c|c|c|c|}
\cline{3-6}
\multicolumn{2}{c|}{} & N-Findr & ULMM & Heylen & UPPNMM\\
\hline
\multirow{3}{*}{$I_1$} & $\Vmat{1}$ & 5.68 & 0.95 &  6.42 & \textbf{\blue{0.27}}\\
\cline{2-6}
                       & $\Vmat{2}$ & 5.85 & \textbf{\blue{0.32}} & 7.46 & 0.36\\
\cline{2-6}
                       & $\Vmat{3}$ & 3.31 & 0.30 & 5.26 & \textbf{\blue{0.27}}\\
\hline
\multirow{3}{*}{$I_2$} & $\Vmat{1}$ & 9.27 & 9.68 & 6.71 & \textbf{\blue{0.59}}\\
\cline{2-6}
                       & $\Vmat{2}$ & 8.58 & 8.67 & 11.80 & \textbf{\blue{0.38}}\\
\cline{2-6}
                       & $\Vmat{3}$ & 4.47 & 6.34 & 4.98 & \textbf{\blue{0.26}}\\
\hline
\multirow{3}{*}{$I_3$} & $\Vmat{1}$ & 7.35 &3.42& 6.48 & \textbf{\blue{1.50}}\\
\cline{2-6}
                       & $\Vmat{2}$ & 10.68 &\textbf{\blue{3.13}} & 11.88 & 3.22\\
\cline{2-6}
                       & $\Vmat{3}$ & 4.34 &7.44& 3.20 & \textbf{\blue{0.85}}\\
\hline
\end{tabular}
\end{center}
\end{footnotesize}
\vspace{-0.4cm}
\end{table}

\begin{table}[h!]
\renewcommand{\arraystretch}{1.2}
\begin{footnotesize}
\begin{center}
\caption{REs ($\times 10^{-2}$): synthetic
images.\label{tab:RE_synth}}
\begin{tabular}{|c|c|c|c|c|}
\cline{3-5}
\multicolumn{2}{c|}{}  &  $I_1$  & $I_2$  &  $I_3$  \\
\multicolumn{2}{c|}{}  &  (LMM)  & (PPNMM)&  (GBM)  \\
\hline
\multirow{2}*{LMM}  & SLMM & $1.04$ & $1.74$& $15.16$\\
 \cline{2-5}
   &  ULMM & $\textbf{\blue{0.99}}$& $1.43 $& $1.07$ \\
\hline
 \multirow{2}*{PPNMM}  & SPPNMM & $1.26 $& $1.27 $ & $1.31 $\\
\cline{2-5}
   &  UPPNMM & $\textbf{\blue{0.99}}$ & $\textbf{\blue{0.99}}$& $\textbf{\blue{0.99}}$\\
\hline
\multicolumn{2}{|c|}{GBM}& $1.27$& $1.64$& $1.33$\\
\hline
\end{tabular}
\end{center}
\end{footnotesize}
\vspace{-0.4cm}
\end{table}
\subsection{Analysis of the estimated nonlinearity parameters} 
\begin{figure}[h!]
  \centering
  \includegraphics[width=\columnwidth]{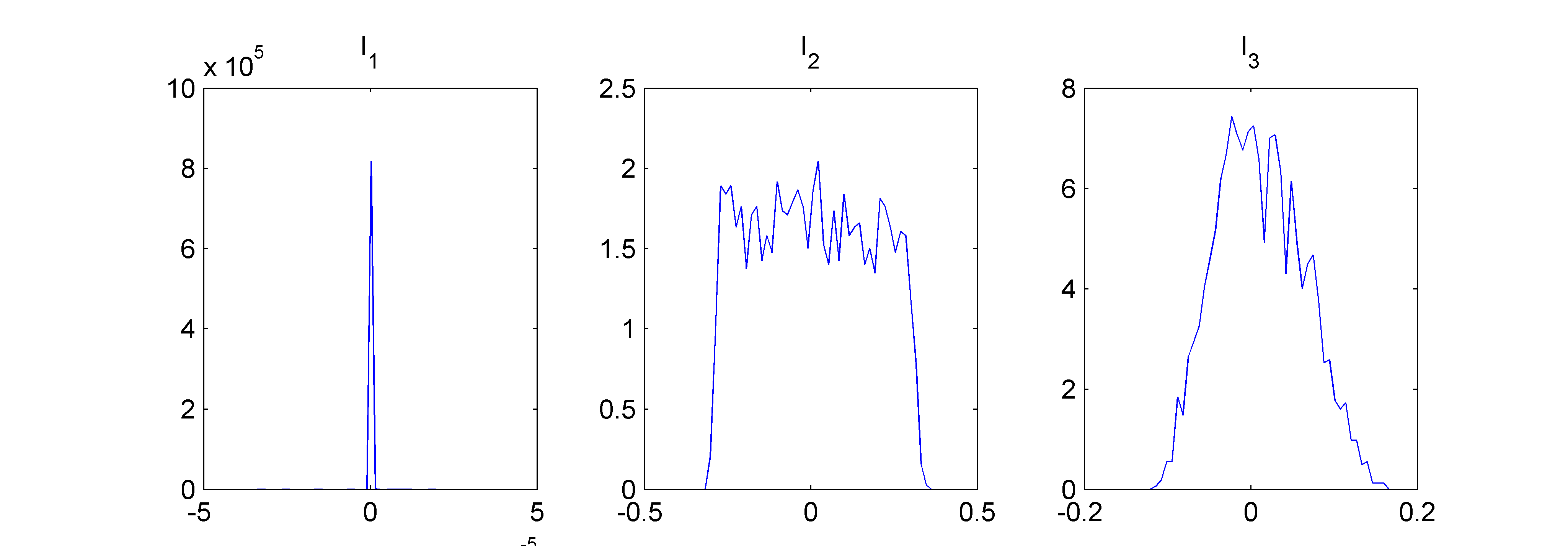}
  \caption{Distributions of the nonlinearity parameters $\nonlin{n}$ for the images $I_1$ (left), $I_2$ (middle) and $I_3$ (right).}
  \label{fig:maps_b_synth}
\end{figure}

As
mentioned above, one of the major properties of the PPNMM is its
ability to characterize the linearity/nonlinearity of the underlying
mixing model for each pixel of the image via the nonlinearity
parameter $\nonlin{n}$. Fig. \ref{fig:maps_b_synth} shows the
nonlinearity parameter distribution estimated for the three images
$I_1$ to $I_3$ using the UPPNMM. This figure shows that the UPPNMM
clearly identifies the linear mixtures of the image $I_1$ whereas
more nonlinearly mixed pixels can be identified in the images $I_2$
and $I_3$. The analysis of Fig. \ref{fig:maps_b_synth} also shows
that the nonlinearities contained in the image $I_3$ (GBM) are
generally less significant than the nonlinearities affecting $I_2$
(PPNMM) for a same signal-to-noise ratio ($\mathrm{SNR} \simeq
21$dB).
\subsection{Performance for different numbers of endmembers} 
\begin{figure}[h!]
  \centering
  \includegraphics[width=\columnwidth]{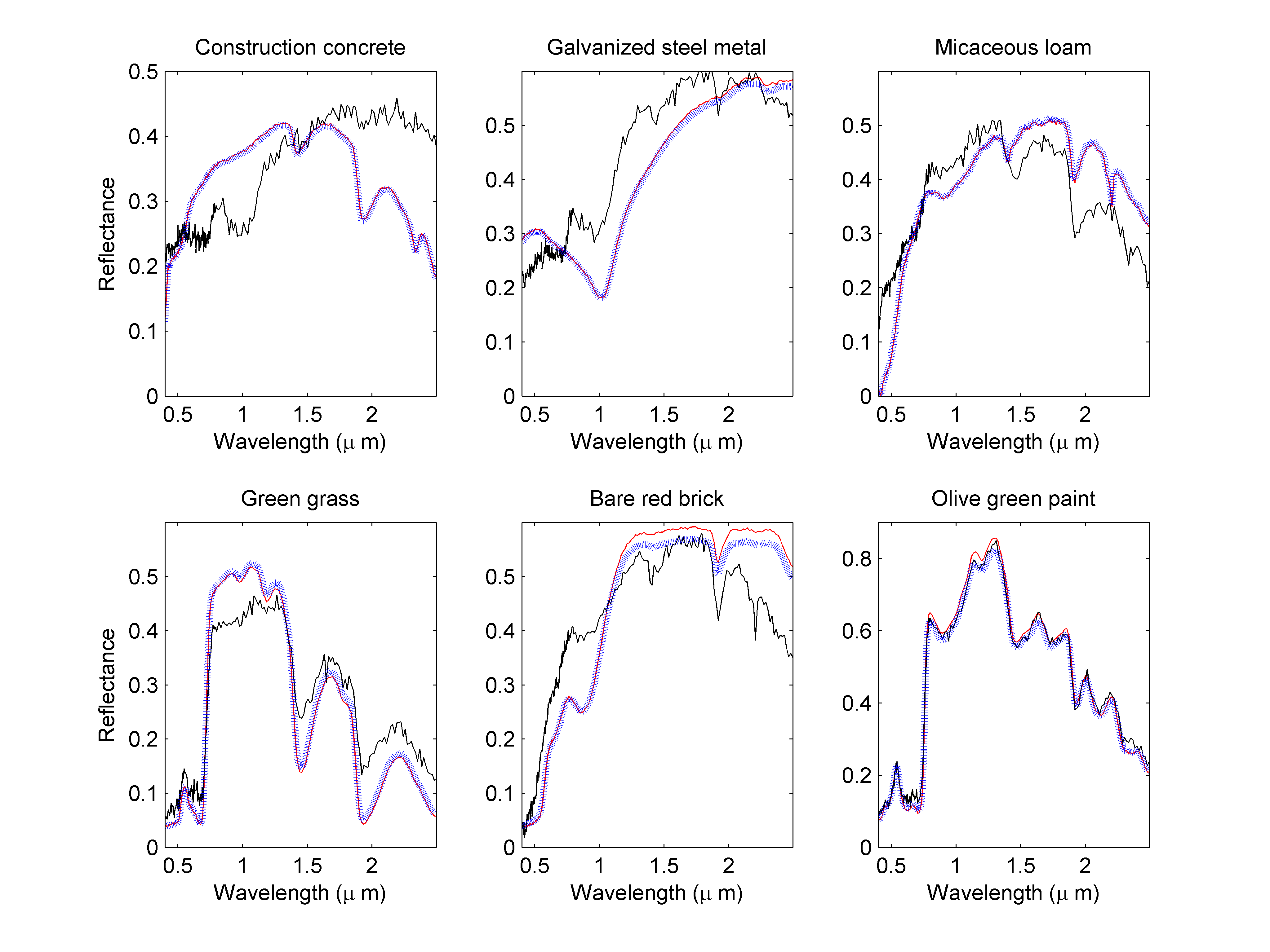}
  \caption{Actual endmembers (blue dots) and the endmembers estimated by Heylen's method (black lines) and the UPPNMM (red lines) for the synthetic image containing
  $R=6$ endmembers.}
  \label{fig:spectres_R6}
\end{figure}

The
next set of simulations analyzes the performance of the proposed
UPPNMM algorithm for different numbers of endmembers ($\nbmat \in
\left \lbrace4,5,6 \right \rbrace$) by unmixing three synthetic
images of $N=2500$ pixels distributed according to the PPNMM. The
endmembers contained in these images have been extracted from the
spectral libraries provided with the ENVI software
\cite{ENVImanual2003}. For each image, the abundance vectors
$\Vabond{n},n=1,\ldots,N$ have been randomly generated according to
a uniform distribution over the admissible set
\eqref{eq:truncated_simplex}. All images have been corrupted by an
additive white Gaussian noise corresponding to $\sigma^2=10^{-4}$.
The nonlinearity coefficients $\nonlin{n}$ are uniformly drawn in
the set $[-0.3,0.3]$. Tables \ref{tab:synth_R} compares the
performance of the proposed method in term of endmember estimation
(average SAMs of the $R$ endmembers), abundance estimation and
reconstruction error. These results show a general degradation of
the abundance and endmember estimations when $R$ is increasing (this
is intuitive since estimator variances usually increase with the
number of parameters to be estimated). However, this degradation is
reasonable when compared to Heylen's method. The proposed algorithm
still provides accurate estimates, as illustrated in Fig.
\ref{fig:spectres_R6} which compares the actual and estimated
endmembers associated with the image containing $R=6$ endmembers.

\begin{table}[h!]
\renewcommand{\arraystretch}{1.2}
\begin{footnotesize}
\begin{center}
\caption{Unmixing performance:synthetic images.\label{tab:synth_R}}
\begin{tabular}{|c|c|c|c|c|}
\cline{3-5}
\multicolumn{2}{c|}{}    & $R=4$  &  $R=5$ & $R=6$ \\
\hline
 \multirow{2}*{Average SAMs ($\times 10^{-2}$)} & SPPNMM    & $7.76$ &$ 10.78$ & $18.53$\\
 \cline{2-5}
 & UPPNMM   &$\textbf{\blue{0.47}}$ & $\textbf{\blue{0.81}}$ & $\textbf{\blue{1.09}}$\\
\hline
 \multirow{2}*{RNMSEs ($\times 10^{-2}$)}   & SPPNMM    & $7.58$ & $10.95$ & $16.52$\\
 \cline{2-5}
  & UPPNMM   & $\textbf{\blue{0.78}}$ & $\textbf{\blue{1.23}}$ & $\textbf{\blue{1.47}}$\\
\hline
 \multirow{2}*{REs ($\times 10^{-2}$)}   & SPPNMM    & $1.36$ & $1.46$ & $1.64$\\
 \cline{2-5}
  & UPPNMM   & $\textbf{\blue{0.99}}$ & $\textbf{\blue{0.99}}$ & $\textbf{\blue{0.99}}$\\
\hline
\end{tabular}
\end{center}
\end{footnotesize}
\vspace{-0.4cm}
\end{table}

\section{Simulations on real data} \label{sec:simu_real}
\subsection{Data sets}
\begin{figure}[h!]
  \centering
  \includegraphics[width=\columnwidth]{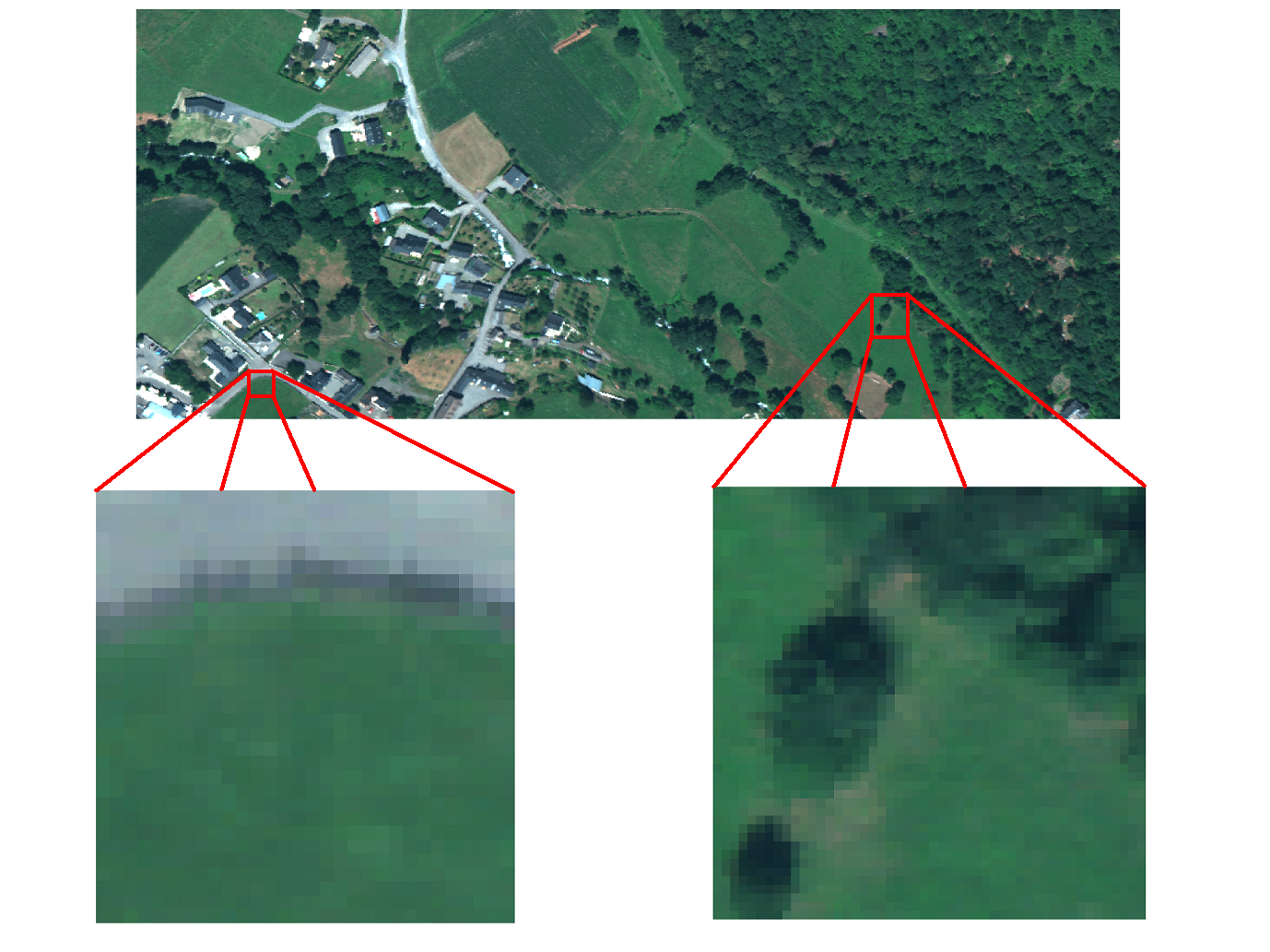}
  \caption{Top: real hyperspectral Madonna data acquired by the Hyspex hyperspectral scanner
   over Villelongue, France. Bottom: Scene $\#1$ (left) and Scene $\#2$ (right) shown in true colors.}
\label{fig:Madonna_big}
\end{figure}

The real image considered in this section was acquired in 2010 by
the Hyspex hyperspectral scanner over Villelongue, France (00° 03'W
and 42°57'N). $L = 160$ spectral bands were recorded from the
visible to near infrared with a spatial resolution of $0.5$m. This
dataset has already been studied in \cite{Sheeren2011,Altmann2013}
and is mainly composed of forested and urban areas. More details
about the data acquisition and pre-processing steps are available in
\cite{Sheeren2011}. Two sub-images denoted as scene $\#1$ and scene
$\#2$ (of size $31 \times 30$ and $50 \times 50$ pixels) are chosen
here to evaluate the proposed unmixing procedure and are depicted in
Fig. \ref{fig:Madonna_big} (bottom images). The scene $\#1$ is
mainly composed of road, ditch and grass pixels. The scene $\#2$ is
more complex since it includes shadowed pixels. For this image,
shadow is considered as an additional endmember, resulting in $R=4$
endmembers, i.e., tree, grass, soil and shadow.
\subsection{Endmember and abundance estimation}
\begin{figure}[h!]
  \centering
  \includegraphics[width=\columnwidth]{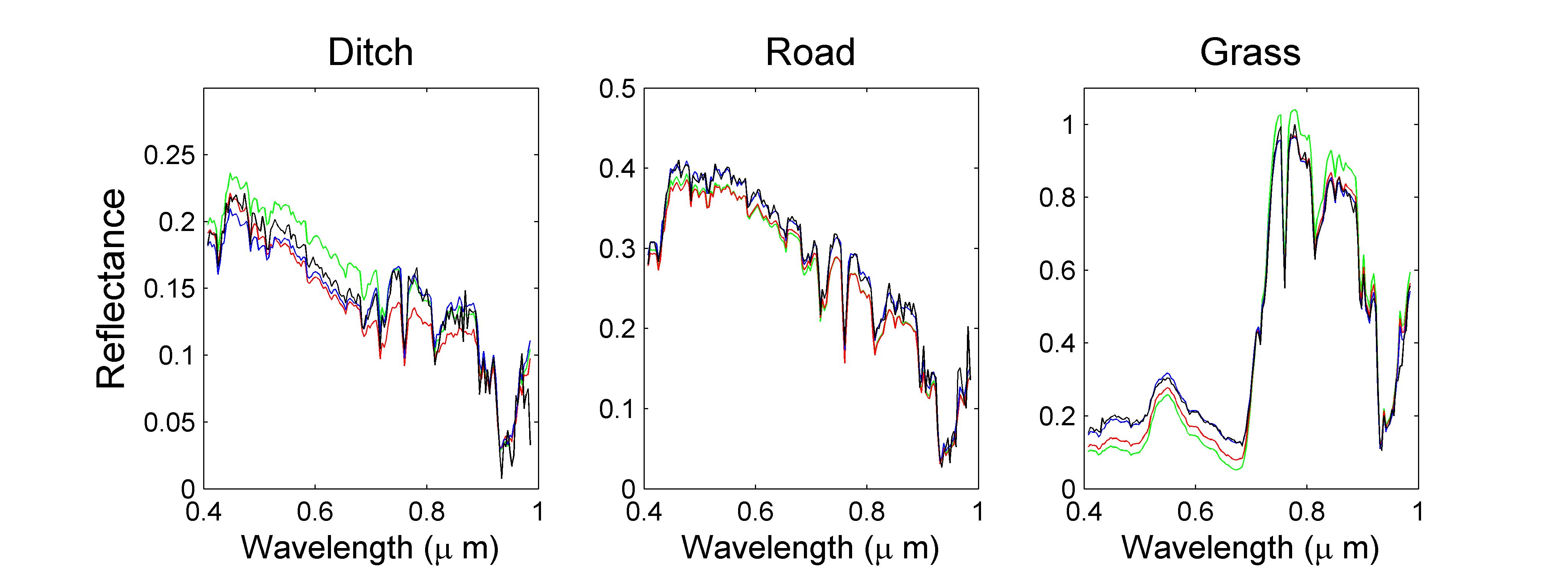}
  \caption{The $R=3$ endmembers estimated by N-Findr (blue lines), ULMM (green lines), Heylen's method (black lines) and the UPPNMM (red lines) for the scene $\#1$.}
  \label{fig:spectres_madonna1}
\end{figure}

\begin{figure}[h!]
  \centering
  \includegraphics[width=\columnwidth]{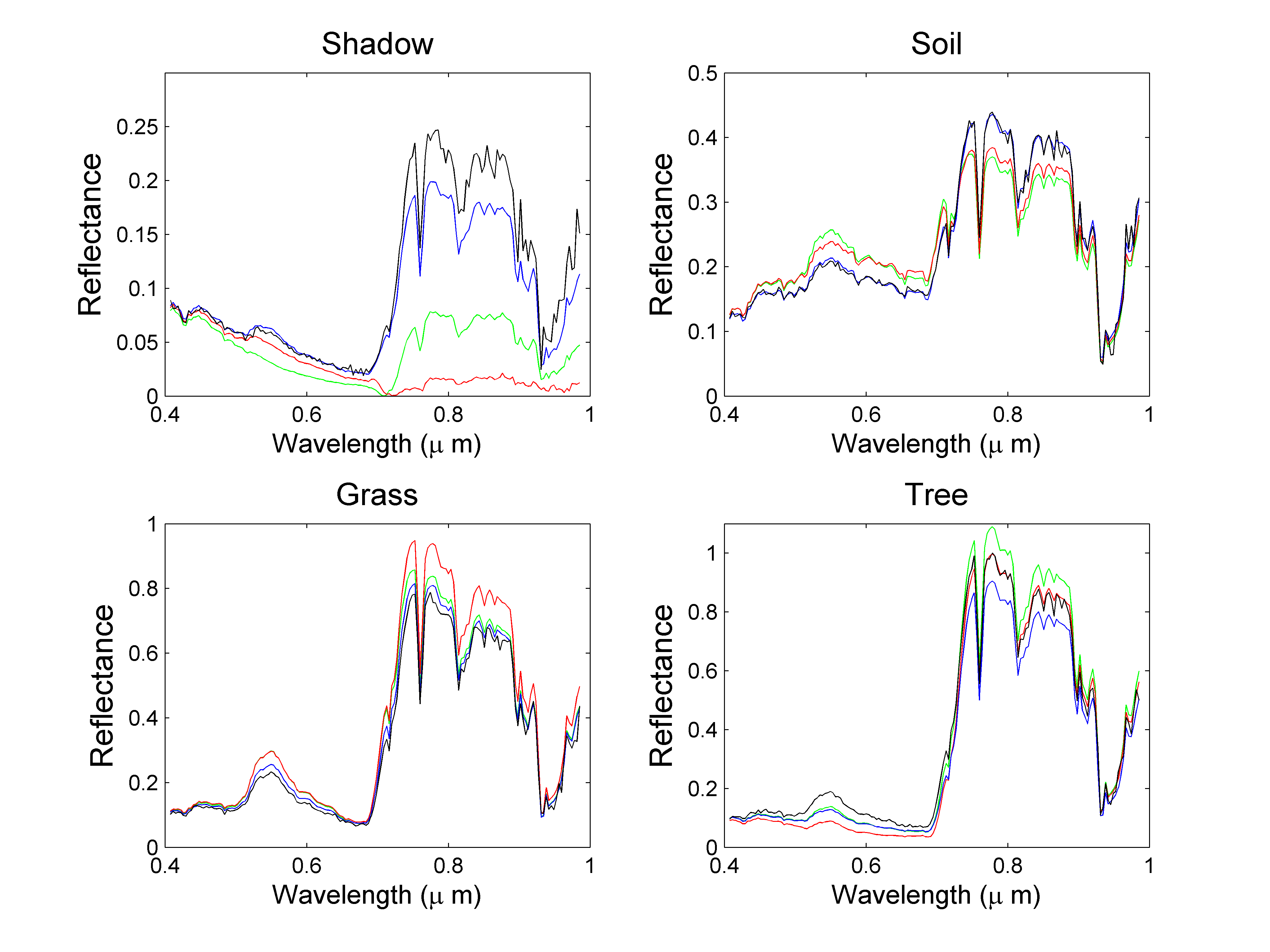}
  \caption{The $R=4$ endmembers estimated by N-Findr (blue lines), ULMM (green lines), Heylen's method (black lines) and the UPPNMM (red lines) for the scene $\#2$.}
  \label{fig:spectres_madonna2}
\end{figure}

\begin{figure}[h!]
  \centering
  \includegraphics[width=\columnwidth]{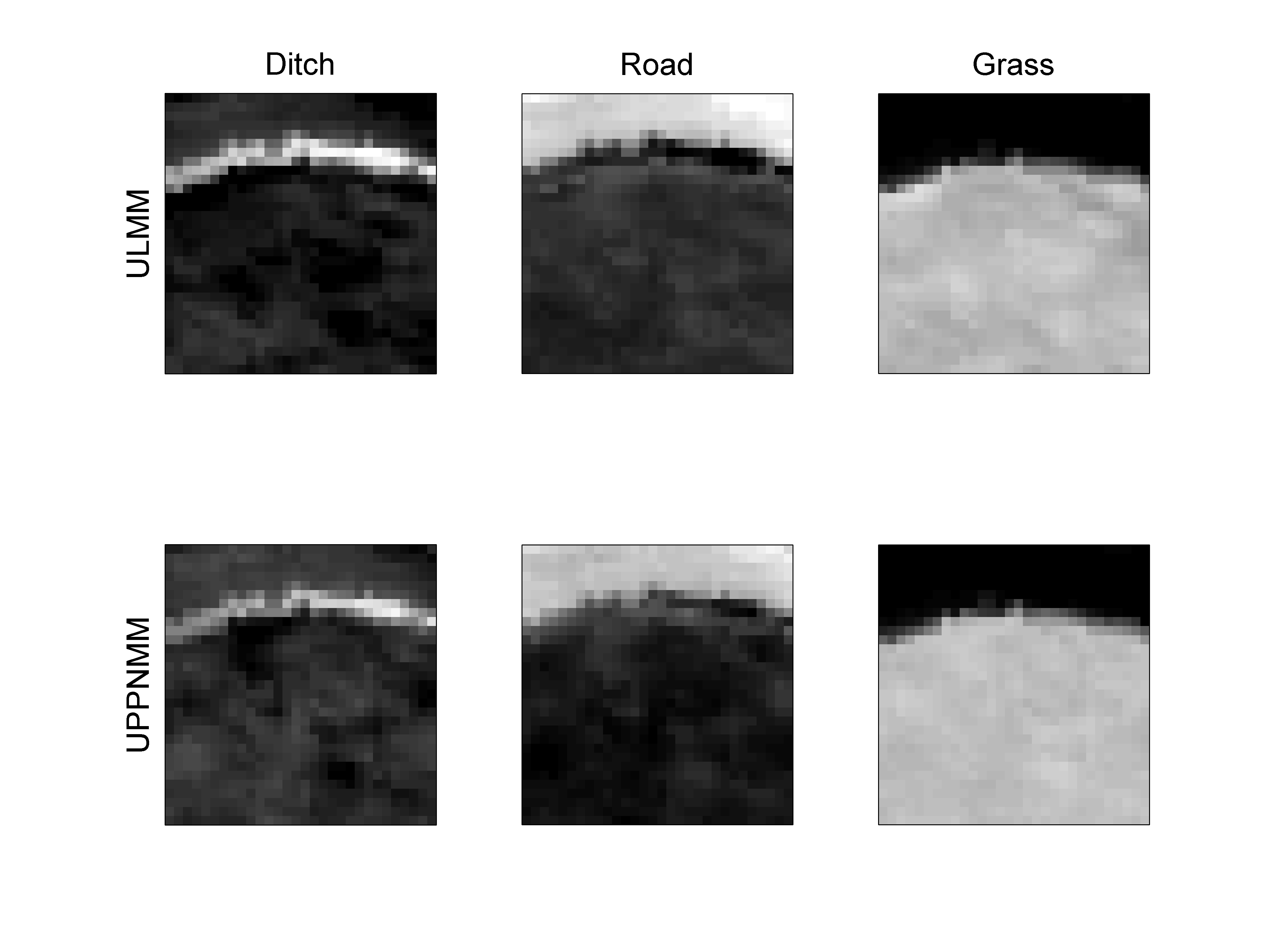}
  \caption{Abundance maps estimated by the SLMM, the GBM and the UPPNMM algorithms for the scene $\#1$.}
  \label{fig:abond_maps_madonna1}
\end{figure}

\begin{figure}[h!]
  \centering
  \includegraphics[width=\columnwidth]{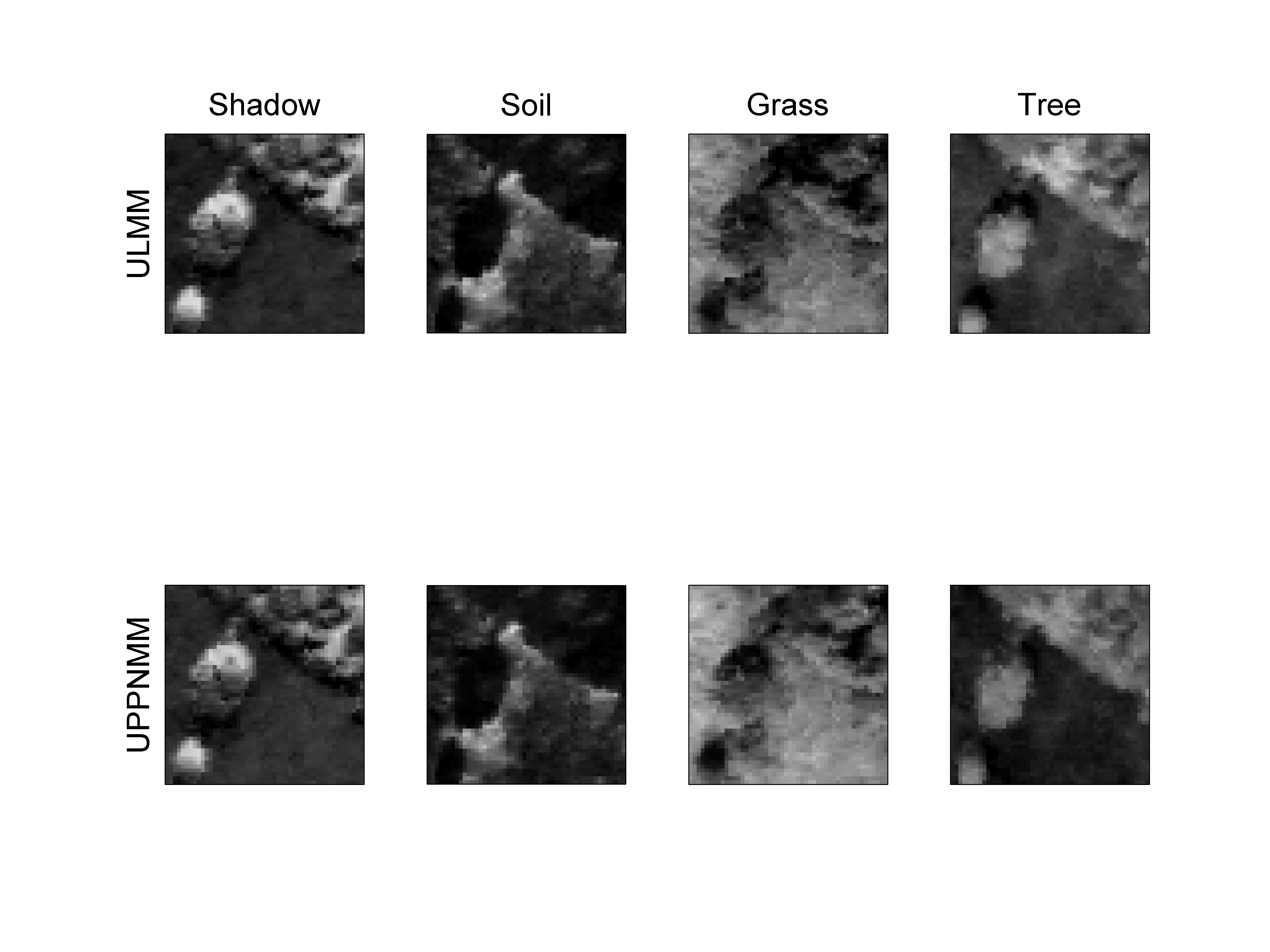}
  \caption{Abundance maps estimated by the SLMM, the GBM and the UPPNMM algorithms for the scene $\#2$.}
  \label{fig:abond_maps_madonna2}
\end{figure}

The endmembers extracted by N-FINDR, the ULMM algorithm
\cite{Dobigeon2009} and Heylen's method \cite{Heylen2011} with $R =
3$ (resp. $R=4$) for the scene $\#1$ (resp. scene $\#2$) are
compared with the endmembers estimated by the UPPNMM in Fig.
\ref{fig:spectres_madonna1} (resp. Fig.
\ref{fig:spectres_madonna2}). For the scene $\#1$, the four
algorithms provide similar endmember estimates whereas the estimated
shadow spectra are different for the scene $\#2$. The N-FINDR
algorithm and Heylen's method estimate endmembers as the purest
pixels of the observed image, which can be problematic when there is
no pure pixel in the image (as it occurs with shadowed pixels in the
scene $\#2$). Conversely, the ULMM and UPPNMM methods, which jointly
estimate the endmembers and the abundances seem to provide more
relevant shadow spectra (of lower amplitude). Examples of abundance
maps for the scene $\#1$ (resp. scene $\#2$), estimated by the ULMM
and the UPPNMM algorithms are presented in Fig.
\ref{fig:abond_maps_madonna1} (resp. Fig.
\ref{fig:abond_maps_madonna2}). The abundance maps obtained by the
UPPNMM are similar to the abundance maps obtained with ULMM.
\subsection{Analysis of nonlinearities}
\begin{figure}[h!]
\begin{minipage}[b]{.48\linewidth}
 \centering
\includegraphics[width=4.8cm]{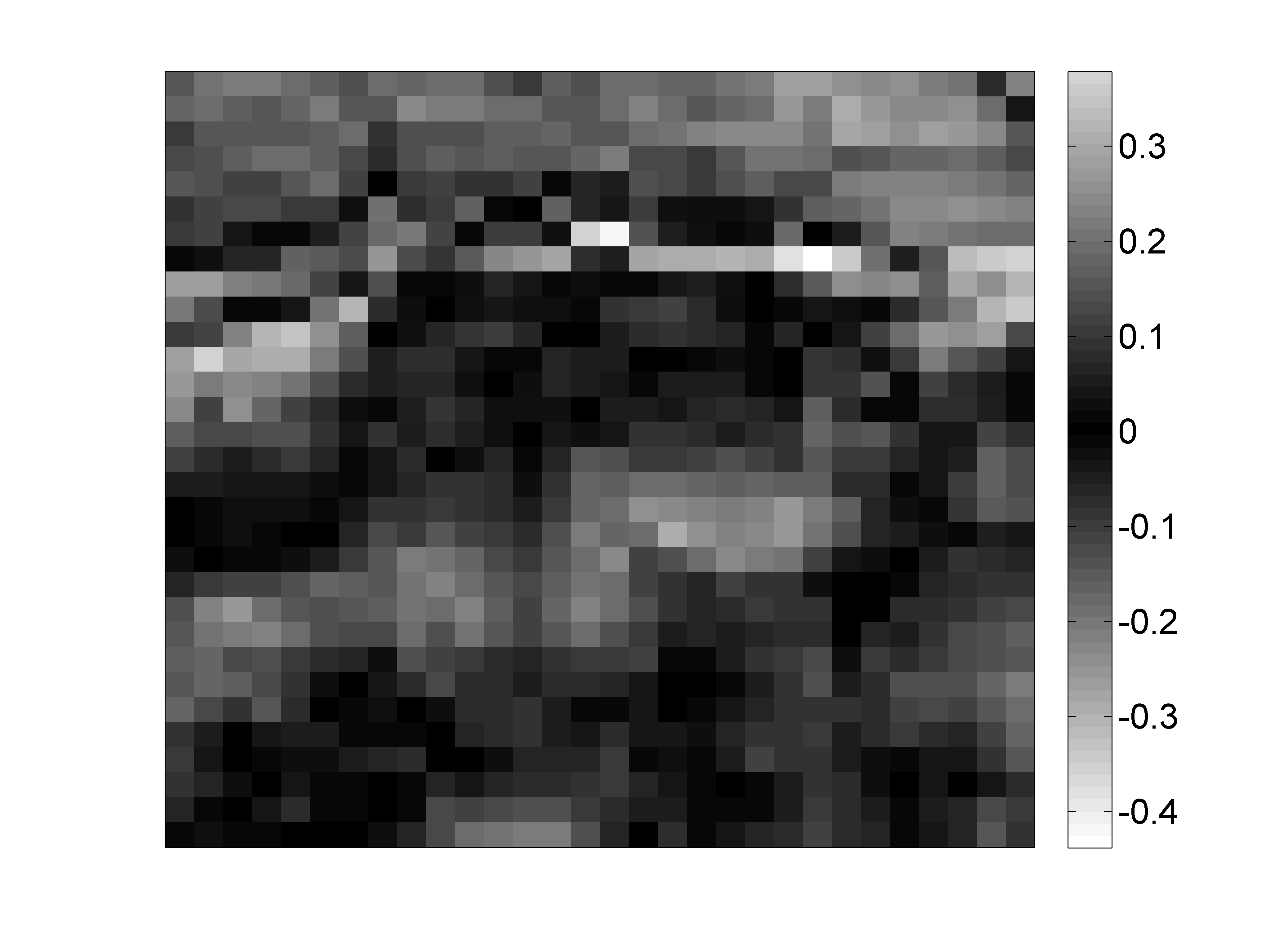}
 \centerline{(a) Scene $\#1$}\medskip
\end{minipage}
\hfill
\begin{minipage}[b]{0.48\linewidth}
 \centering
\includegraphics[width=4.8cm]{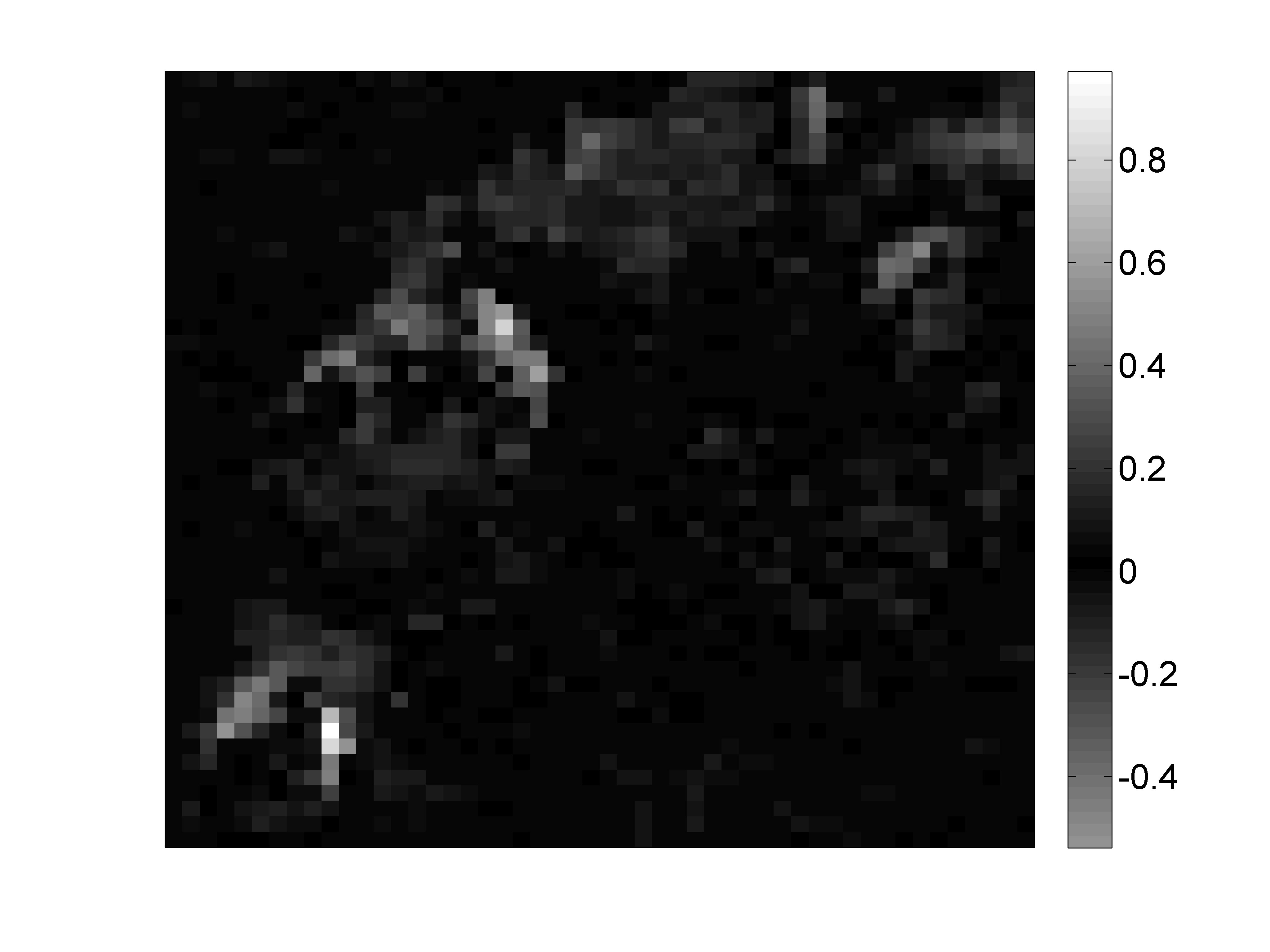}
 \centerline{(b) Scene $\#2$}\medskip
\end{minipage}
\vspace{-0.3cm} \caption{Maps of the nonlinearity parameter
$\nonlin{n}$ estimated by the UPPNMM for the real images.}
\label{fig:b_maps_madonna}
\end{figure}

Fig. \ref{fig:b_maps_madonna} shows the estimated maps of
$\nonlin{n}$ for the two considered images. Different nonlinear
regions can be identified in the scene $\#1$, mainly in the
grass-planted region (probably due to endmember variability) and
near the ditch (presence of relief). For the scene $\#2$, nonlinear
effects are mainly detected in shadowed pixels.
\subsection{Estimation of noise variances}
\begin{figure}[h!]
  \centering
  \includegraphics[width=\columnwidth]{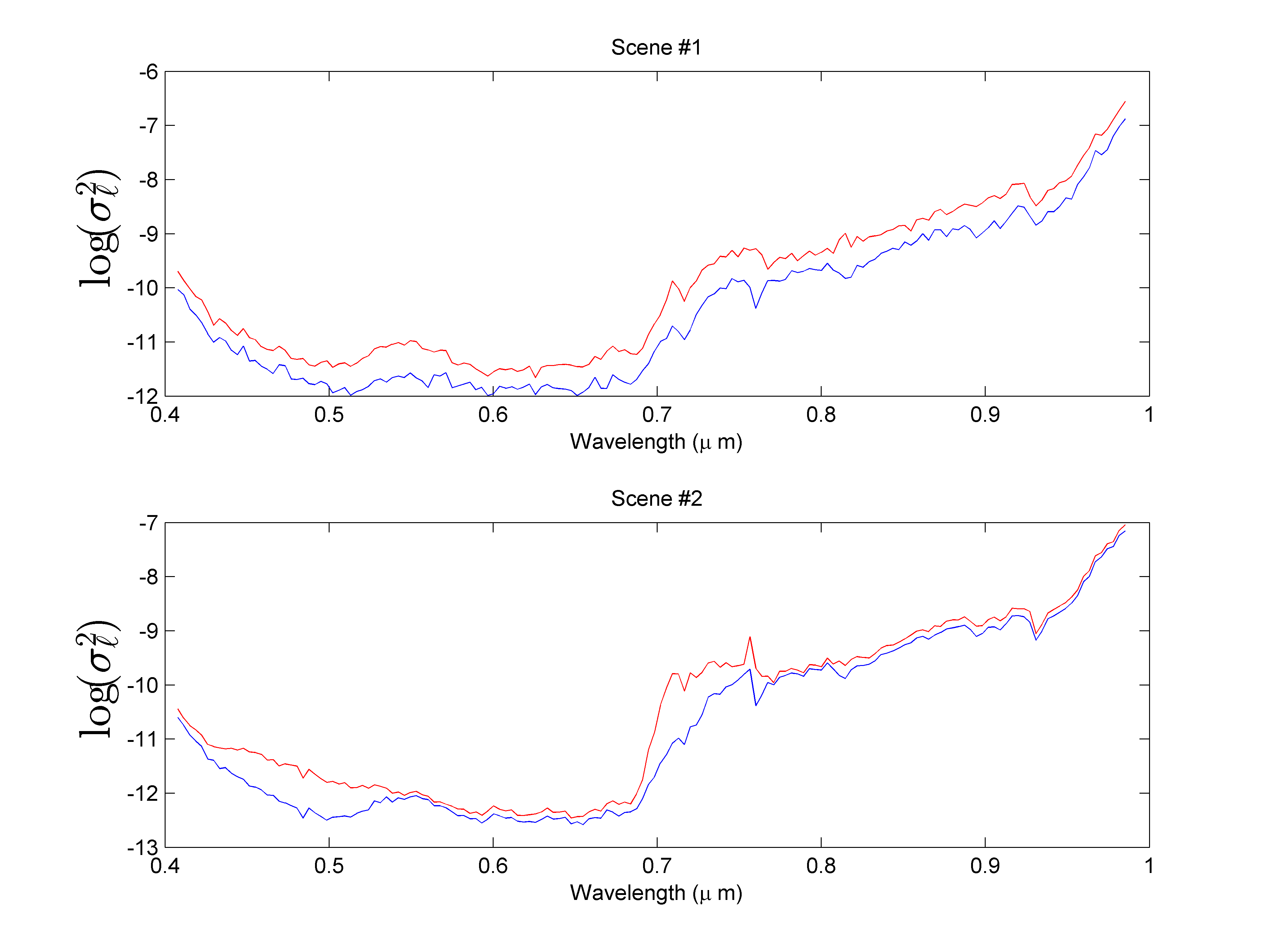}
  \caption{Noise variances estimated by the UPPNMM (red) and the Hysime algorithm (blue) for the scene $\#1$ (top) and the scene $\#2$ (bottom).}
  \label{fig:noise variances_madonna}
\end{figure}

Fig. \ref{fig:noise variances_madonna} compares the noise variance
estimated by the UPPNMM for the two real images with the noise
variance estimated by the HySime algorithm \cite{Bioucas2008}. The
HySime algorithm assumes additive noise and estimates the noise
covariance matrix of the image using multiple regression. Fig.
\ref{fig:noise variances_madonna} first shows that the two
algorithms provides similar noise variance estimates. Moreover,
these results motivate the consideration of non i.i.d. noise for
hyperspectral image analysis since the noise variances increase for
the higher wavelengths for the two images.

\subsection{Image reconstruction}
The proposed algorithm is finally evaluated from the REs associated
with the two real images. These REs are compared in Table
\ref{tab:RE_real} with those obtained by assuming other mixing
models.  The two unsupervised algorithms (ULMM and UPPNMM) provide
smaller REs than the SU procedures decomposed into two steps. This
observation motivates the use of joint abundance and endmember
estimation algorithms.
\begin{table}[h!]
\renewcommand{\arraystretch}{1.2}
\begin{footnotesize}
\begin{center}
\caption{REs ($\times 10^{-2}$): Real image.\label{tab:RE_real}}
\begin{tabular}{|c|c|c|c|}
 \cline{3-4}
\multicolumn{2}{c|}{}  &  Scene $\#1$  & Scene $\#2$ \\
\hline
\multirow{2}*{LMM}  & SLMM & $1.53$ & $1.04$\\
 \cline{2-4}
   &  ULMM & $1.11$ & $\textbf{\blue{0.88}}$\\
\hline
 \multirow{2}*{PPNMM}  & SPPNMM & $1.50$ & $1.17$\\
\cline{2-4}
   &  UPPNMM & $\textbf{\blue{1.08}}$ & $0.89$\\
\hline
\multicolumn{2}{|c|}{GBM}& $1.72$ & $1.25$ \\
\hline
\end{tabular}
\end{center}
\end{footnotesize}
\vspace{-0.4cm}
\end{table}

\section{Conclusions and future work}
\label{sec:conclusion} We proposed a new hierarchical Bayesian
algorithm for unsupervised nonlinear spectral unmixing of
hyperspectral images. This algorithm assumed that each pixel of the
image is a post-nonlinear mixture of the endmembers contaminated by
additive Gaussian noise. The physical constraints for the abundances
and endmembers were included in the Bayesian framework through
appropriate prior distributions. Due to the complexity of the
resulting joint posterior distribution, a Markov chain Monte Carlo
method was used to approximate the MMSE estimator of the unknown
model parameters. Because of the large number of parameters to be
estimated, Hamiltonian Monte Carlo methods were used to reduce the
sampling procedure complexity and to improve the mixing properties
of the proposed sampler. Simulations conducted on synthetic data
illustrated the performance of the proposed algorithm for linear and
nonlinear spectral unmixing. An important advantage of the proposed
algorithm is its flexibility regarding the absence of pure pixels in
the image. Another interesting property resulting from the
post-nonlinear mixing model is the possibility of detecting
nonlinearly from linearly mixed pixels. This detection can identify
the image regions affected by nonlinearities in order to
characterize the nonlinear effects more deeply. The number of
endmembers contained in the hyperspectral image was assumed to be
known in this work. We think that estimating the number of
components present in the image is an important issue that should be
considered in future work. Finally, considering endmember
variability in linear and nonlinear mixing models is an interesting
prospect which is currently under investigation.

\section*{Appendix: Derivation of the potential functions}

\label{sec:appendix} The potential energy \eqref{eq:potential_z} can
be rewritten
\begin{eqnarray}
U(\Vtabond{n})&  = & U_1(\Vabond{n}) + U_2(\Vtabond{n})
\end{eqnarray}
where
\begin{eqnarray}
 U_1(\Vabond{n}) & = & \dfrac{1}{2}  \left[\Vpix{n}-\bg_n\left(\MATmat\Vabond{n}\right)\right]\transp\bSigma^{-1} \left[\Vpix{n}-\bg_n\left(\MATmat\Vabond{n}\right)\right],\nonumber\\
 U_2(\Vtabond{n}) & = & -\sum_{r=1}^{R-1}
\log\left(\tabond{r}{n}^{\nbmat-r-1}\right).\nonumber
\end{eqnarray}
Partial derivatives of $U(\Vtabond{n})$ with respect to
$\Vtabond{n}$ is obtained using the classical chain rule
\begin{eqnarray}
\dfrac{\partial U(\Vtabond{n})}{\partial \Vtabond{n}} =
\dfrac{\partial U_1(\Vabond{n})}{\partial \Vabond{n}}
\dfrac{\partial \Vabond{n}}{\partial \Vtabond{n}} + \dfrac{\partial
U_2(\Vtabond{n})}{\partial \Vtabond{n}}\nonumber
\end{eqnarray}
Straightforward computations lead to\\

$\dfrac{\partial U_1(\Vabond{n})}{\partial \Vabond{n}} =$
\begin{eqnarray}
 - \left[\Vpix{n}-\bg_n\left(\MATmat\Vabond{n}\right)\right]\transp\bSigma^{-1}
\left[\MATmat+
2\nonlin{n}\left(\MATmat\Vabond{n}\Vun{R}\transp\right)\odot\MATmat\right]\nonumber
\end{eqnarray}
\begin{eqnarray}
\dfrac{\partial \abond{r}{n}}{\partial \tabond{i}{n}} & = & \left\{
    \begin{array}{lll}
        0 & \mbox{if } &i>r \\
        \dfrac{\abond{r}{n}}{\tabond{i}{n}-1} & \mbox{if }& i=r \\
        \dfrac{\abond{r}{n}}{\tabond{i}{n}} & \mbox{if }& i<r \\
    \end{array}
\right.\nonumber\\
\dfrac{\partial U_2(\Vtabond{n})}{\partial \tabond{i}{n}} & = &
-\dfrac{\nbmat-i-1}{\tabond{i}{n}}.
\end{eqnarray}
Similarly, the potential energy \eqref{eq:potential_m} can be
rewritten
\begin{eqnarray}
V(\Vmat{\ell,:})&  = & V_1(\btt_{\ell}) + V_2(\Vtabond{n})
\end{eqnarray}
with $\btt_{\ell}=\MATabond\transp \Vmat{\ell,:} +
\diag{\Vnonlin}\left[\left(\MATabond\transp \Vmat{\ell,:} \right)
\odot \left(\MATabond\transp \Vmat{\ell,:} \right) \right]$ and
\begin{eqnarray}
V_1(\btt_{\ell}) & = & \dfrac{\Vert \Vpix{\ell,:}- \btt_{\ell} \Vert
^2}{2\sigma_{\ell}^2}\nonumber\\
V_2(\Vmat{\ell,:}) & = & \dfrac{\Vert \Vmat{\ell,:}-
\bar{\bm}_{\ell,:} \Vert ^2}{2s^2}.\nonumber
\end{eqnarray}

The partial derivatives of the potential energy
\eqref{eq:potential_m} can be obtained using the chain rule
\begin{eqnarray}
\dfrac{\partial V(\Vmat{\ell,:})}{\partial \Vmat{\ell,:}} =
\dfrac{\partial V_1(\btt_{\ell})}{\partial \btt_{\ell}}
\dfrac{\partial \btt_{\ell}}{\partial \Vmat{\ell,:}} +
\dfrac{\partial V_2(\Vmat{\ell,:})}{\partial\Vmat{\ell,:}}\nonumber
\end{eqnarray}
and
\begin{eqnarray}
\dfrac{\partial V_1(\btt_{\ell})}{\partial \btt_{\ell}} & = & -\dfrac{ (\Vpix{\ell,:}- \btt_{\ell})\transp}{\sigma_{\ell}^2}\nonumber\\
\dfrac{\partial \btt_{\ell}}{\partial \Vmat{\ell,:}} & = & \MATabond\transp + 2\diag{\Vnonlin}\left[\left(\MATabond\transp\Vmat{\ell,:}\Vun{R}\transp\right)\odot\MATabond\transp\right]\nonumber\\
\dfrac{\partial V_2(\Vmat{\ell,:})}{\partial\Vmat{\ell,:}} & = &
\dfrac{ (\Vmat{\ell,:}- \bar{\bm}_{\ell,:})\transp}{s^2}\nonumber
\end{eqnarray}

\bibliographystyle{IEEEtran}
\bibliography{biblio}

\end{document}